\documentclass[12pt]{article}
\usepackage{epsfig}

\newsavebox{\LSIM}
\sbox{\LSIM}{\raisebox{-1ex}{$\ \stackrel{\textstyle<}{\sim}\ $}}

\newsavebox{\GSIM}
\sbox{\GSIM}{\raisebox{-1ex}{$\ \stackrel{\textstyle>}{\sim}\ $}}

\newcommand{\be} {\begin{equation}}
\newcommand{\ee} {\end{equation}}
\newcommand{\bdm} {\begin{displaymath}}
\newcommand{\edm} {\end{displaymath}}
\newcommand{\bc} {\begin{center}}
\newcommand{\ec} {\end{center}}
\newcommand{\beqa} {\begin{eqnarray}}
\newcommand{\eeqa} {\end{eqnarray}}
\newcommand{\nn} {\nonumber}

\newcommand{\bear}{\begin{eqnarray}}
\newcommand{\ear}{\end{eqnarray}}
\newcommand{\bea}{\begin{eqnarray*}}
\newcommand{\ea}{\end{eqnarray*}}
\newcommand{\rf}{\ref}
\newcommand{\lb}{\label}

\newcommand{\slp}{\raise.15ex\hbox{$/$}\kern-.57em\hbox{$\partial$}}
\newcommand{\slG}{\raise.15ex\hbox{$/$}\kern-.57em\hbox{$G$}}
\newcommand{\slA}{\raise.15ex\hbox{$/$}\kern-.57em\hbox{$A$}}
\newcommand{\grgl}{\:\hbox to -0.2pt{\lower2.5pt\hbox{$\sim$}\hss}
{\raise3pt\hbox{$>$}}\:}
\newcommand{\klgl}{\:\hbox to -0.2pt{\lower2.5pt\hbox{$\sim$}\hss}
{\raise3pt\hbox{$<$}}\:}

\newcommand{\befi}[1]{\begin{figure}[ht] \leavevmode \centering \epsffile{#1.eps}}

\newcommand{\beq}{\begin{equation}}
\newcommand{\enq}{\end{equation}}
\newcommand{\beqast}{\begin{eqnarray*}}
\newcommand{\enqa}{\end{eqnarray}}
\newcommand{\enqast}{\end{eqnarray*}}

\newcommand{\al}{\alpha}
\newcommand{\ga}{\gamma}

\newcommand{\ep}{\epsilon}

\newcommand{\si}{\sigma}

\newcommand{\ph}{\phi}

\def\ct{\cite}
\def\pmb#1{\setbox0=\hbox{#1}
\kern.05em\copy0\kern-\wd0 \kern-.025em\raise.0433em\box0 }

\def\ell{l}

\def\ln{\log}

\begin{document}

\begin{flushright} HD THEP-01-27\\
 M/C-TH-01/04
\end{flushright}
\bc
{\bf\Large A Comprehensive Approach to Structure Functions}
\ec
\vskip 1truecm
\bc
A. Donnachie\\
Department of Physics and Astronomy,
University of Manchester\\Manchester M13 9PL, UK\\
email: {\tt ad@a3.ph.man.ac.uk}
\ec
\bc
H.G.Dosch\\
Institut f\"ur Theoretische Physik der Universit\"at Heidelberg\\
Philosophenweg 16, D-69120 Heidelberg\\
email: {\tt h.g.dosch@thphys.uni-heidelberg.de}
\ec
\vskip 2truecm
\bc
{\bf Abstract}
\ec
\medskip
We present a model based on a dipole picture with a hard and a soft pomeron
in which large dipoles couple to the soft pomeron and small dipoles couple 
to the hard pomeron. The parameters in the model are fixed by proton-proton
scattering and the proton structure function $F_2(x,Q^2)$. The model is then
applied successfully to the proton charm structure function $F_2^c(x,Q^2)$, 
the proton longitudinal structure function $F_2^L(x,Q^2)$, $J/\psi$ 
photoproduction, deep virtual Compton scattering $\gamma^* p \to \gamma p$, 
the real photon-proton total cross section $\sigma^{\rm Tot}_{\ga p}(s)$,
the real photon-photon total cross section $\sigma^{\rm Tot}_{\ga\ga}(s)$, 
and the photon structure function $F_2^\gamma(x,Q^2)$. Differences between our 
predictions and and data on charm production in real photon-photon 
interactions and the $\ga^*\ga^*$ cross section $\sigma^{\rm Tot}_{\ga\ga}(s)$
are discussed. 
\newpage

\section{Introduction}

The suggestion \cite{DL98,DL01} that deep inelastic scattering at small $x$ 
can be economically and successfully described by a two-component model 
comprising the soft nonperturbative pomeron of hadronic interactions, 
with intercept $\sim 1.08$, and a hard pomeron, with intercept 
$\approx 1.4$ has met with considerable
phenomenological success when applied to other reactions.  Notable among these
are $J/\psi$ photoproduction and the charm structure function of the proton
\cite{DL99,DL01}, and exclusive $\rho$ and $\phi$ photoproduction at large $t$
\cite{DL00}. 
Successful although this phenomenology is, it does not explain,
for example, the relative strengths of the hard and soft pomeron in deep
inelastic scattering or in $J/\psi$ photoproduction; or why the charm structure
function of the proton is completely dominated by  the hard pomeron. 

To answer questions like these requires a specific model for the diffractive
process. This in turn necessitates consideration of the particle wave functions
which enter the reactions, and to disentangle the dynamics of diffraction
from wave-function effects it is necessary to treat several processes 
simultaneously. An example is provided by \cite{DGS01} in which high-energy
exclusive photo- and electroproduction of vector mesons were studied in a
two-component model of diffraction. The soft and hard pomerons were modelled
by nonperturbative and perturbative gluon exchange respectively. This approach
has the advantage of providing a common kinematical structure in which it is
possible to separate the effects of the vector-meson wave functions from the
dynamics of the exchange. It was shown that the wave functions determine
many aspects of the data, including some which might have been considered
to reflect the dynamics of the exchange.

In this paper we follow the same philosophy and treat hadron-hadron, 
photon-hadron, and photon-photon reactions in a uniform approach, but
with a different model for the two pomerons. The model is based 
\cite{Rue99,DDR99,DDR00} on a dipole 
picture with two pomerons in which small dipoles couple to the 
hard pomeron and large dipoles to the soft pomeron. The proton is considered
as a quark-diquark system i.e. effectively as a dipole. This is very
convenient but not essential for the approach~\cite{DFK94}. The dipole-dipole 
cross section \ct{KD91,DFK94} has been obtained in a functional approach 
\ct{Nac91} to high-energy hadron-hadron scattering, the functional integrals 
being approximately evaluated in a specific nonperturbative model, the 
stochastic vacuum model \ct{Dos87,DS88}.
This model yields confinement and  relates high energy scattering with low
energy data and with results of lattice gauge calculations. The total
dipole-dipole cross  section is obtained as the forward scattering amplitude of
two dipoles  averaged over all orientations. This is then transportable to any 
dipole-dipole-type reaction for which the wave functions of the participating
particles are known.

In section 2 we quote the results required for
the present calculation and refer to the literature for motivation and
justification. In section 3 the model is applied in turn to the proton
structure function $F_2(x,Q^2)$, the proton charm structure function
$F_2^c(x,Q^2)$, the proton longitudinal structure function $F_2^L(x,Q^2)$,
$J/\psi$ photoproduction, deep virtual Compton scattering $\gamma^* p \to
\gamma p$, the real photon-proton total cross section 
$\sigma_{\ga p}^{\rm Tot}(W^2)$, the real photon-photon total cross section 
$\sigma_{\ga\ga}^{\rm Tot}(W^2)$, 
the photon structure function $F_2^\gamma(x,Q^2)$, charm production in 
real photon-photon interactions, and the virtual photon-photon cross
section $\sigma_{\ga^*\ga^*}^{\rm Tot}(W^2)$. The parameters for the 
dipole-dipole cross 
section are fine-tuned to proton-proton  scattering and the criteria 
for defining small and large 
dipoles are obtained from the proton structure function. These 
parameters remain unchanged throughout,
and all other processes are controlled by the relevant particle wave 
functions. Our conclusions are presented in section 4.

\section{The model}

Our normalisation of the forward scattering amplitude $T_{ab \to cd}$ for the 
reaction $ab \to cd$ is such that the forward elastic cross section is 
given by:
\beq
\frac{d}{dt}\si_{ab \to cb}\Big|_{t=t_{\rm min}} = \frac{1}{16 \pi^2 W^4} 
|T_{ab \to cb}|^2.
\label{amp}
\enq
If the outgoing particle $c$ is the same as the incoming particle
$a$ we obtain the total cross section from the optical theorem as
\beq
\si^{\rm Tot}_{ab} = \frac{1}{W^2} {\rm Im}T_{ab\to ab},
\label{opt}
\enq 
where $W$ is the centre-of-mass energy of particles $a$ and $b$.

In the model one calculates the expectation value of two light-like 
Wilson loops with
transverse extension $\vec R_1$ and $\vec R_2$. After averaging over all
directions and integrating over the  impact parameter  one
obtains the forward scattering amplitude of two dipoles \cite{KD91,DFK94}.
The dipole-dipole cross section is obtained using the optical theorem 
(\ref{opt}). At $W = \sqrt{s} = 20$ GeV
it can be numerically  approximated to an accuracy of better than $10 \%$ 
by the factorising form \beq
\sigma_{\rm dip} (R_1,R_2) =  
0.67 \frac{1}{4 \pi^2}(\langle g^2FF\rangle a^4)^2\;
R_1\Big(1-e^{-\frac{R_1}{3.1 a}}\Big)\,
R_2\Big(1-e^{-\frac{R_2}{3.1 a}}\Big)\,
\lb{sidip}
\enq
where $\langle g^2FF\rangle$ is the gluon condensate in a pure
gauge theory  and $a$ is the correlation length of the gauge-invariant 
two-gluon
correlator.  The parameters are taken from lattice results \ct{DDM97} and
fine-tuned to  $pp$ scattering:
\beq
a= 0.346 \mbox{ fm } \qquad \langle g^2FF\rangle a^4 = 23.77.
\enq
A quark-diquark picture is used for the proton so that the dipole formalism
is applicable.

Then $T_{a b \to c b}$ is obtained by 
multiplying (\ref{sidip}) with the products of the appropriate wave 
functions, using (\ref{opt}) and integrating:
\beqa
T_{ab\to cb} &=&i W^2 \int d^2 R_1 \,d^2R_2\,\int_0^1dz_1\,dz_2
\psi_c^*(\vec R_1,z_1)\psi_a(\vec R_1,z_1) |\psi_b(\vec R_2,z_2)|^2
\nn \\
&&~~~~~~~\times \sigma_{\rm dip}(R_1,R_2)~. \enqa
Here $\vec{R}$ is the 
distance vector from the quark to the antiquark or diquark and $z$ is the 
longitudinal momentum fraction of the quark. It is assumed that the product
$\psi_c^*(\vec R_1,z_1)\psi_a(\vec R_1,z_1)$ depends at most weakly on the 
polar angle $\ph$ of $\vec R$ and so the $\phi$-dependence can be ignored.

For the photon wave functions we use the perturbative expressions:
\beqa 
\psi_\ga^{\lambda=0, h,\bar h}(Q^2,\vec{R},z)&=& 
\sqrt{3\alpha} \hat{e}_f \frac{- 2 z (1-z) Q \delta_{h,-\bar h}}{2\pi} 
Q\,K_0(\ep R)\nn \\ 
\psi_\ga^{\lambda=\pm1, h,\bar h}(Q^2,\vec{R},z)&=& 
\sqrt{3 \alpha} \hat{e}_f 
\frac{\pm \sqrt{2}}{2 \pi} \bigg( i\, e^{\pm i \ph} 
(z \delta_{h,\frac{1}{2}}\delta_{\bar h,-\frac{1}{2}}\nn \\ 
&&\hspace{-2cm} 
-(1- z) \delta_{h,-\frac{1}{2}}\delta_{\bar h,\frac{1}{2}})\,K_1(\ep R) 
+m\,  \delta_{h,\pm \frac{1}{2}}\delta_{\bar h,\pm \frac{1}{2}}K_0(\ep R) 
\bigg) 
\label{phowa} 
\enqa  
with 
\beq
\ep=\sqrt{z(1-z)Q^2 +m_f^2}~.
\lb{epsilon}
\enq
Here $R$ and $\ph$ are the plane-polar coordinates of the transverse 
separation $\vec{R}$ of the  quark-antiquark pair, $\hat e_f$ is the charge 
of the quark in units of the elementary charge,
$m_f$ its  mass, and $h$ and $\bar h$ the helicities of the
quarks and the antiquarks; $\lambda=0$  indicates a longitudinal photon,
$\lambda=\pm 1$ a transverse photon. The functions $K_i$ are the modified
Bessel functions. These expressions can be used for photons of high 
virtuality.
For photons of low virtuality we use the same expressions but with a 
$Q^2$-dependent mass $m_{\rm eff}(Q^2)$ instead of $m_f$. This procedure
has been
justified in \ct{DGP98} and the following linear parametrisations 
have been obtained from
comparison with the phenomenological vector-current two-point function:
\beq 
m_{\rm eff}(Q^2)= \left\{ \begin{array} {l@{\qquad \mbox{for} \qquad} l}
m_f+m_{0q}\,(1-Q^2/Q_0^2)& Q^2\le Q_0^2\\
 m_f & Q^2\ge Q_0^2
\end{array} \right.
\label{meff} 
\enq 
with 
\beq
\begin{array}{lll}
m_{0q}=0.20\pm0.02 \mbox{ GeV }& m_f=0.007 \mbox{ GeV } &
Q_0^2=1.05 \mbox{ GeV}^2
\end{array}
\enq  
for the up and down quark, and 
\beq
\begin{array}{lll}
m_{0q}=0.31\pm0.02 \mbox{ GeV }& m_f=0.15 \mbox{ GeV } &
Q_0^2=1.6 \mbox{ GeV}^2
\end{array}
\enq  
for the strange quark. In this paper we use $m_{0q}=0.19$ GeV for the
light quarks and 0.31 GeV for the strange quark. The mass of the
charmed quark was chosen as the median  value of the $\overline{MS}$-mass at
$\mu=m_c$,  $m_c=1.25$ GeV  \cite{PDG00}.

In the quark-diquark picture of the proton we use a Gaussian wave function  
\beq  \psi_p(\vec{R})
=\frac{1}{\sqrt{2 \pi} \,  R_p} \exp\bigg(- \frac{R^2}{4  R_p^2}\bigg) .
\label{wav}  
\enq  
The transverse radius $R_p$ was also fine tuned \cite{DGKP97},
in this case to obtain the observed 
logarithmic slope for elastic $p p$ scattering at a centre-of-mass energy of 
$W=20$ GeV: 
\beq
R_p=0.75 \mbox{ fm}.
\enq

The wave function for the $J/\psi$ is taken from \ct{DGKP97}. It is
constructed in the following way. The spin structure is that of a massive
vector current with mass $m_c$, that is it has the same structure as the 
charm part
of the photon wave function (\rf{phowa}). An additional $z$-dependent factor 
as introduced in \ct{WSB85} is also included and the
dependence on the transverse distance $R$ is modelled by a Gaussian like
(\rf{wav}). The mean radius is fixed by the normalisation condition and the
electromagnetic decay width. 

The stochastic vacuum model is a model for the infrared behaviour of QCD and
and was applied originally to hadron-hadron scattering alone
\ct{DFK94,FP97,BN99}. It turned  out that it yielded reasonable results for
photon-induced processes   for photon virtualities $Q^2$ up to about 10 GeV$^2$
\ct{DGKP97,DGP98, KDP99}. For higher values of $Q^2$ the model overestimates
the cross sections. This may have the following reason. 
For consistency of the model with low-energy theorems the strong coupling in
the infrared domain 
must have the frozen value \ct{RD95} $\alpha_s
\approx 0.57$. It is plausible that upon introduction of a hard scale
through a  highly-virtual photon the coupling of the gluons to the
corresponding  dipole is governed by that hard scale. Therefore we have 
rescaled
the results obtained for the dipole cross section (1) by the factor   
\beq
\frac{\al_s(Q^2)}{\al_s(0)}=\frac{1}{0.57}\cdot \frac{4
\pi}{11 \cdot \log(Q^2/Q_0^2+7.42)} \lb{supp} \enq 
with $Q_0^2=1$ GeV$^2$. This corresponds to a running coupling
$\alpha_s(Q^2)$  in a flavourless scheme adjusted to give $\alpha_s(0)=0.57$.

As mentioned in the Introduction our main purpose is to apply the two-pomeron 
approach
to different processes. The results of \ct{DL98,DL01} strongly suggest 
that the 
soft pomeron couples predominantly to large dipoles whereas the hard pomeron
couples to small dipoles. In order to be economical with parameters
we introduce a sharp cut and assume that only the soft pomeron couples
if both dipoles are larger than a certain value $R_c$, whereas the hard 
pomeron couples if at least one of the dipoles is smaller than $R_c$. Energy
dependence is introduced by hand into the the dipole cross section 
(\ref{sidip}) by dividing the amplitude into a soft and a hard part
with the coefficients $\si_{s}$ and $\si_h$:
\beq
T_{ab \to cb}(W) =i W^2\Big( \si_{s}~(W/W_0)^{2 \ep_s} 
+ \si_h ~(W/W_0)^{2 \ep_h}\Big)
\lb{soha}\enq
with 
\beq
W_0=20 \mbox{ GeV }, ~~ \ep_s=0.08, ~~ \ep_h=0.42~.
\enq

The soft pomeron contribution  is given by:
\beqa
\si_s&=&\int_{R_c}^\infty 2 \pi d R_1 \,\int_{R_c}^\infty 2 \pi d R_2\,
\int_0^1dz_1\,dz_2
\psi_c^*(\vec R_1,z_1)\psi_a(\vec R_1,z_1) |\psi_b(\vec R_2,z_2)|^2 \nn\\
&&~~~~~~~~~~\times \sigma_{\rm dip}(R_1,R_2).
\enqa

For the 
hard pomeron it has been argued \cite{FKS99} that the appropriate 
dimensionless 
variable is $RW$ for the following reason. Highly virtual
photons have a hadronic radius 
$R \propto 1/Q$, so in order to ensure scaling behaviour for the
dimensionless quantity  $Q^2\si_{\rm dip}(R,W)$ the $W$ dependence
should come in the combination $W^2R^2$ which corresponds to the inverse 
of the Bjorken variable $x$. If one dipole is small, say $R_1\leq R_c$,
the hard contribution should depend upon the factor $(R_1 W)$; if both dipoles
are small, then upon the factor $(R_1R_2 W^2)$. Since the factor $W^{2\ep_h}$ 
has been extracted in (\rf{soha}), we obtain  for the coefficient $\si_h$:
\beqa 
\si_h&=&\int^{R_c}_0 2 \pi d R_1 \,\int_{0}^{R_c} 2 \pi d R_2\,
\int_0^1dz_1\,dz_2
\psi_c^*(\vec R_1,z_1)\psi_a(\vec R_1,z_1) |\psi_b(\vec R_2,z_2)|^2\nn\\
&&~~~~~~~~~~ \times \sigma_{\rm dip}(R_1,R_2)
(R_1R_2/R_c^2)^{\ep_h}\nn\\
&&+\int^{R_c}_0 2 \pi d R_1 \,\int_{R_c}^{\infty} 2 \pi d R_2\,
\int_0^1dz_1\,dz_2
\psi_c^*(\vec R_1,z_1)\psi_a(\vec R_1,z_1) |\psi_b(\vec R_2,z_2)|^2\nn\\
&&~~~~~~~~~~ \times \sigma_{\rm dip}(R_1,R_2)
(R_1/R_c)^{2 \ep_h}\nn\\
&&+\int^{R_c}_0 2 \pi d R_2 \,\int_{R_c}^{\infty} 2 \pi d R_1\,
\int_0^1dz_1\,dz_2
\psi_c^*(\vec R_1,z_1)\psi_a(\vec R_1,z_1) |\psi_b(\vec R_2,z_2)|^2 \nn\\
&&~~~~~~~~~~\times \sigma_{\rm dip}(R_1,R_2)
(R_2/R_c)^{2 \ep_h}
\lb{sidipW}
\enqa

\section{Results}

\subsection{$\ga^* p$ reactions}

\begin{figure}
\leavevmode
\begin{center}
\begin{minipage}{54mm}
\epsfxsize54mm
\epsffile{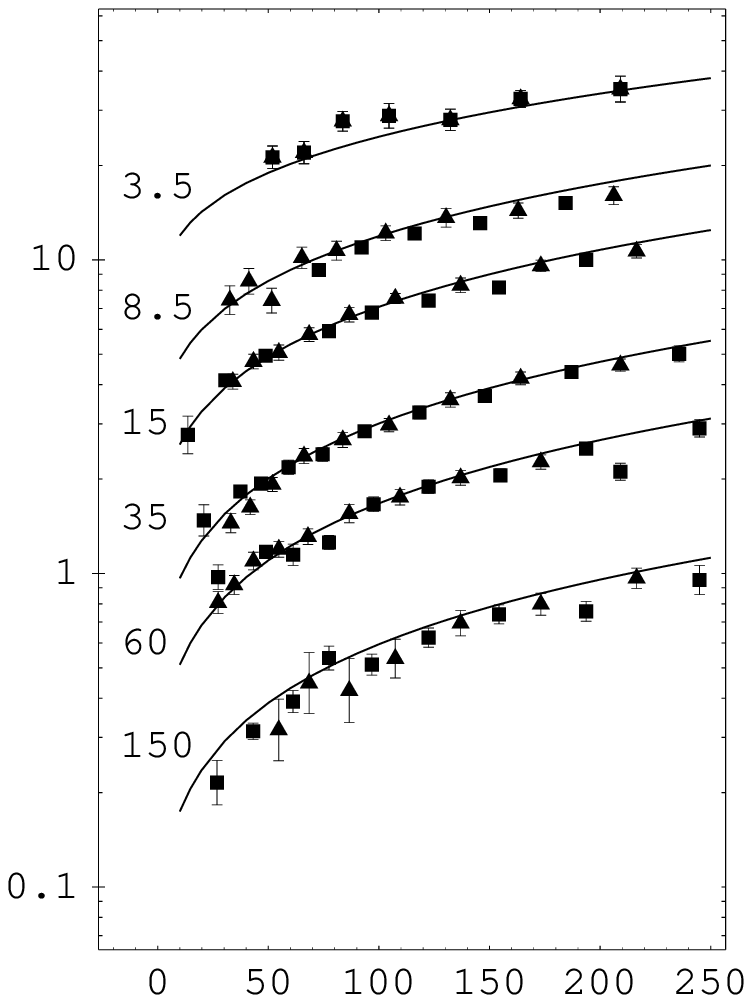}
\begin{picture}(0,0)
\setlength{\unitlength}{.9mm}
\put(30,2){\small{$W$ [GeV]}}
\put(-2,80){$\si_{\ga^*p}$}
\put(0,75){$\mu$b}
\end{picture}
\end{minipage}
\end{center}
\begin{center}
\begin{minipage}{5.4cm}
\epsfxsize5.4cm
\epsffile{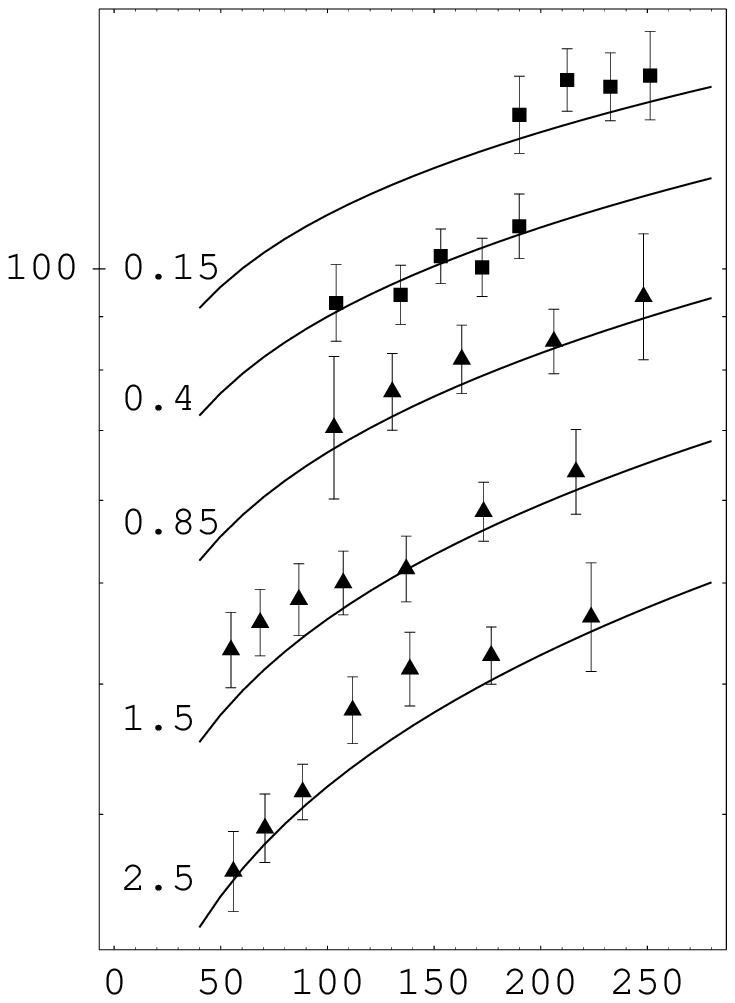}
\begin{picture}(0,0)
\setlength{\unitlength}{.9mm}
\put(30,2){\small{$W$ [GeV]}}
\put(-2,80){$\si_{\ga^*p}$}
\put(0,75){$\mu$b}
\end{picture}
\end{minipage}
\hfill
\begin{minipage}{5.4cm}
\epsfxsize5.4cm
\epsffile{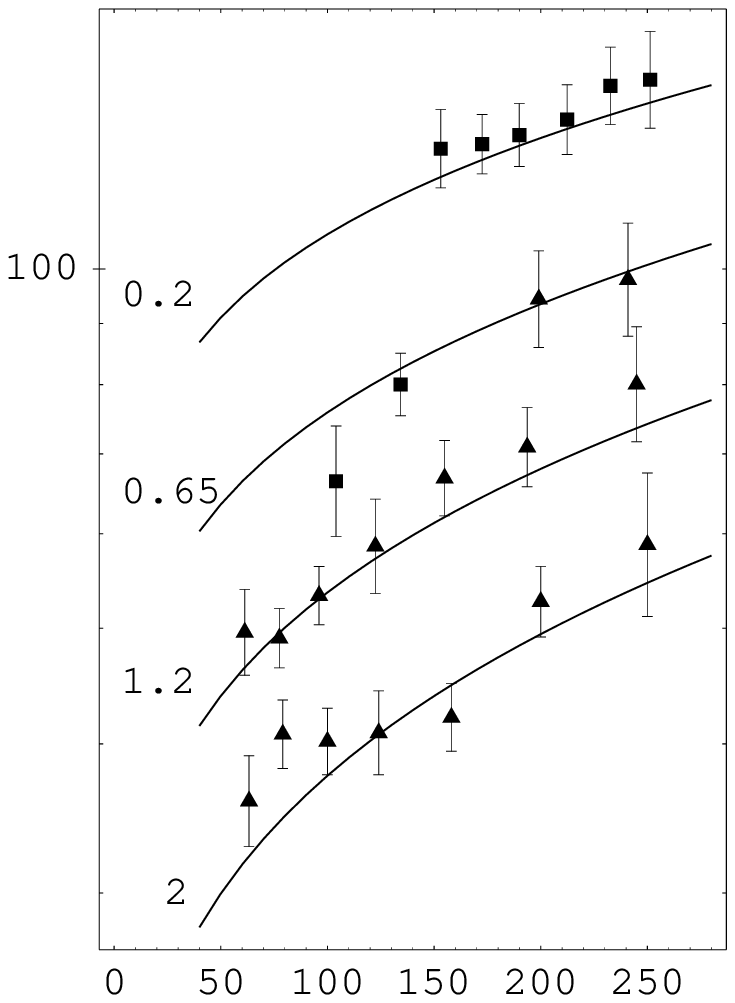}
\begin{picture}(0,0)
\setlength{\unitlength}{.9mm}
\put(30,2){\small{$W$ [GeV]}}
\put(-2,80){$\si_{\ga^*p}$}
\put(0,75){$\mu$b}
\end{picture}
\end{minipage}
\end{center}
\caption{Examples of $\si^{\rm Tot}_{\ga^* p}(Q^2,W)$ for different
values of the photon virtuality $Q^2$ in GeV$^2$ as
indicated in the figures. The solid line is our model. The data
are from ZEUS \protect \ct{ZEUS_X}, squares, and H1
\protect \ct{H1_X}, triangles.} 
\lb{gap} 
\end{figure}
\begin{figure}
\leavevmode
\vspace{1cm}
\begin{center}
\begin{minipage}{54mm}
\epsfxsize54mm
\epsffile{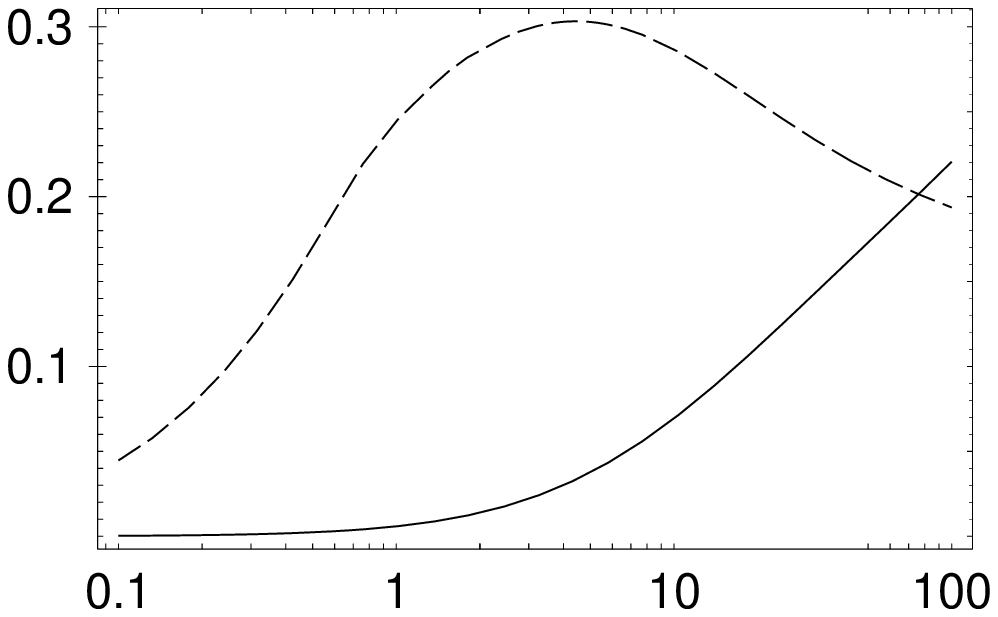}
\begin{picture}(0,0)
\setlength{\unitlength}{1mm}
\put(-10,30){$F_2$}
\put(25,42){$x=0.1$}
\end{picture}
\end{minipage}
\begin{minipage}{51mm}
\epsfxsize51mm
\epsffile{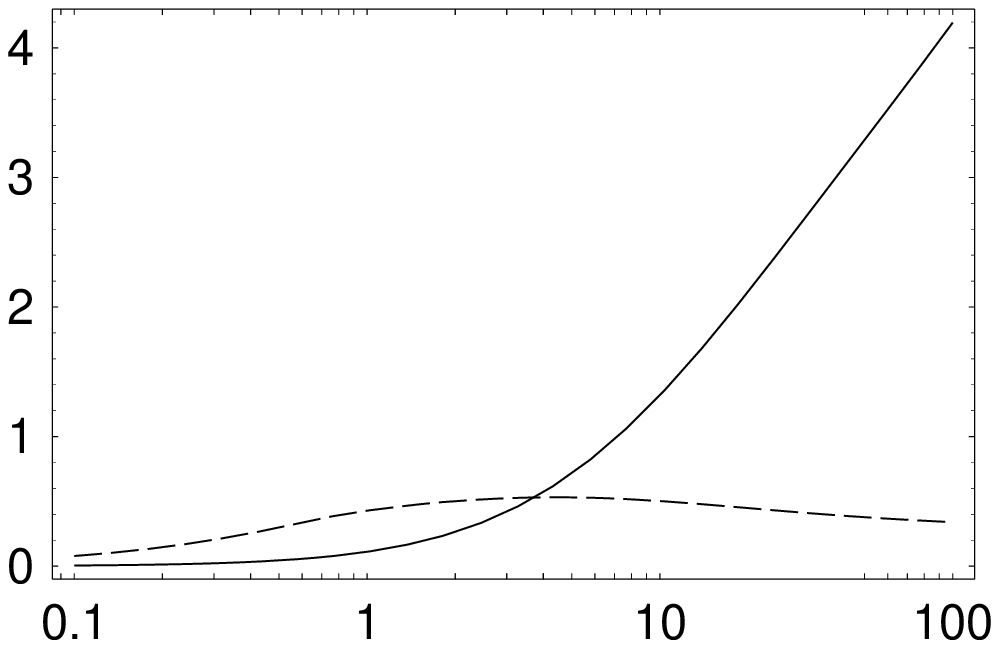}
\begin{picture}(0,0)
\setlength{\unitlength}{1mm}
\put(20,41){$x=0.0001$}
\end{picture}
\end{minipage}
\end{center}
\vspace{-13mm}
\begin{center}
\begin{minipage}{55mm}
\epsfxsize55mm
\epsffile{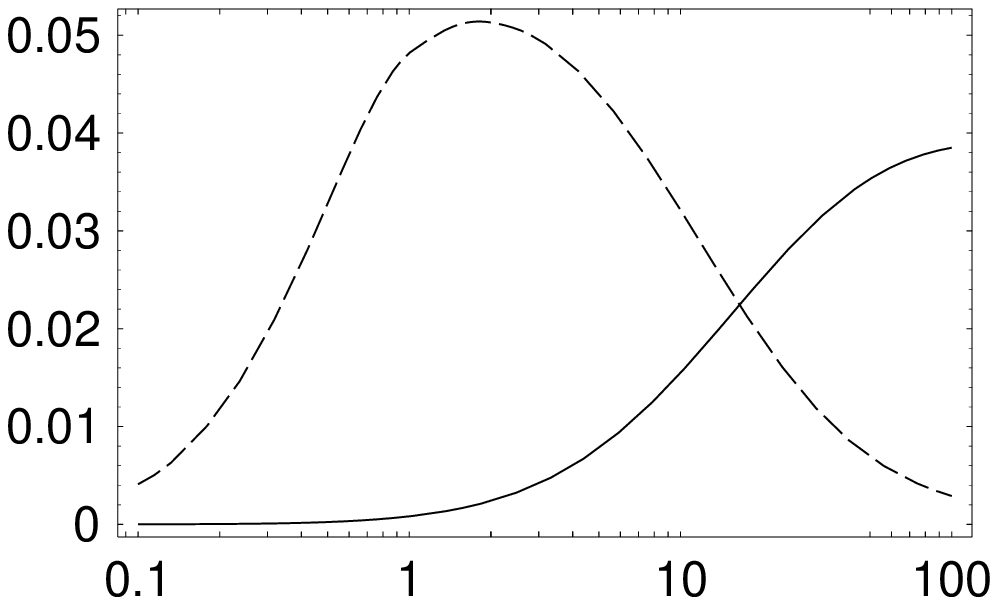}
\begin{picture}(0,0)
\setlength{\unitlength}{1mm}
\put(-10,30){$F_L$}
\end{picture}
\end{minipage}
\begin{minipage}{55mm}
\epsfxsize55mm
\epsffile{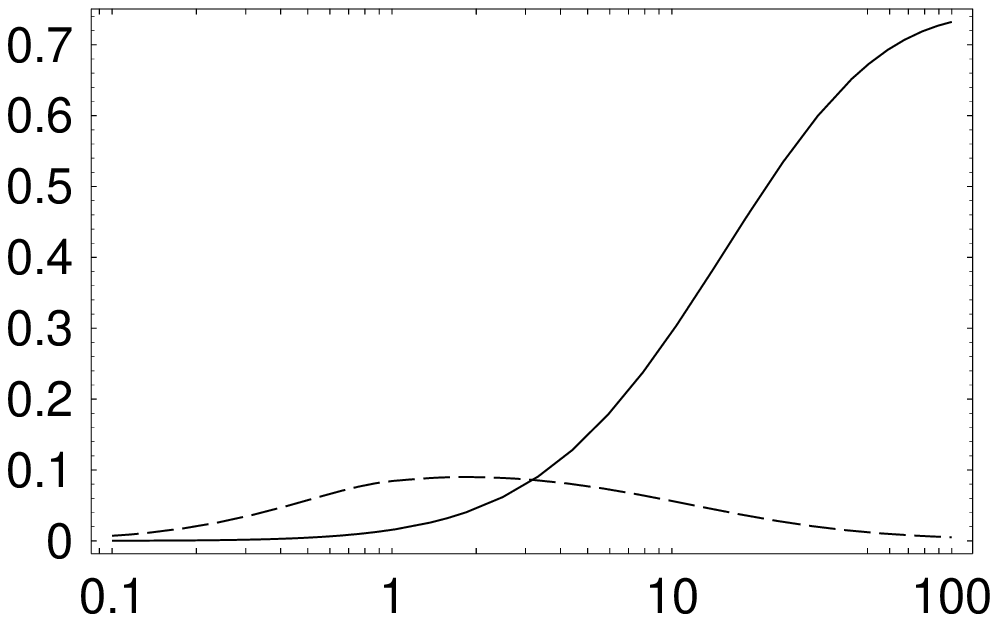}
\begin{picture}(0,0)
\setlength{\unitlength}{1mm}
\end{picture}
\end{minipage}
\end{center}
\vspace{-15mm}
\begin{center}
\begin{minipage}{55mm}
\epsfxsize55mm
\epsffile{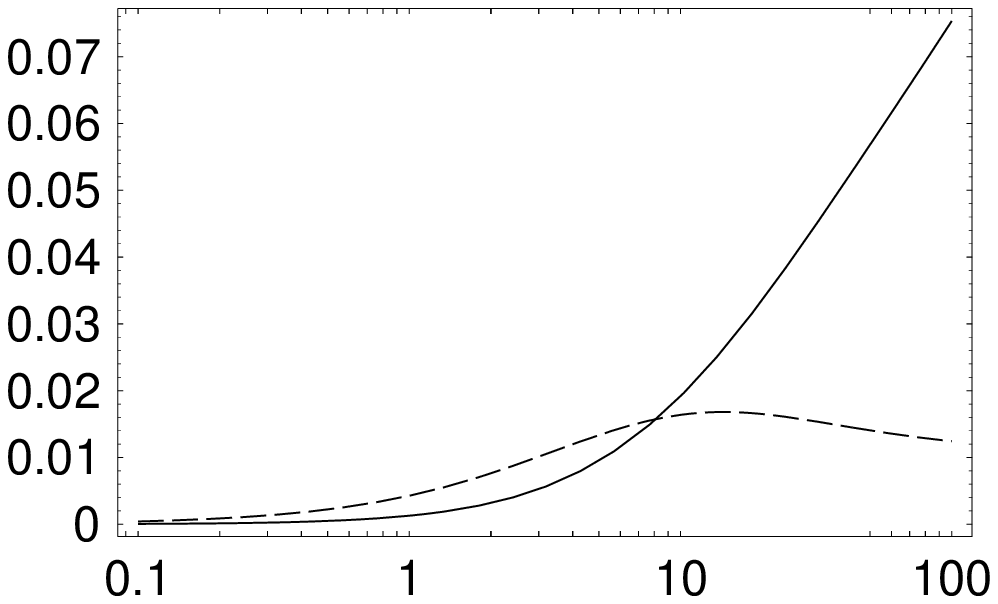}
\begin{picture}(0,0)
\setlength{\unitlength}{1mm}
\put(-10,30){$F^c_2$}
\end{picture}
\end{minipage}
\begin{minipage}{55mm}
\epsfxsize55mm
\epsffile{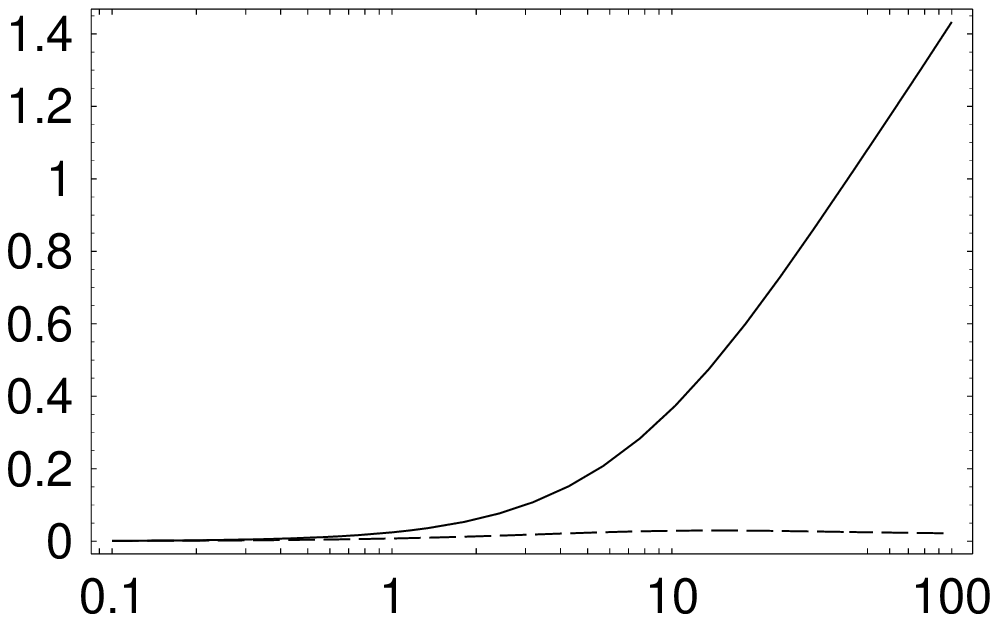}
\begin{picture}(0,0)
\setlength{\unitlength}{1mm}
\end{picture}
\end{minipage}
\end{center}
\vspace{-15mm}
\begin{center}
\begin{minipage}{55mm}
\epsfxsize55mm
\epsffile{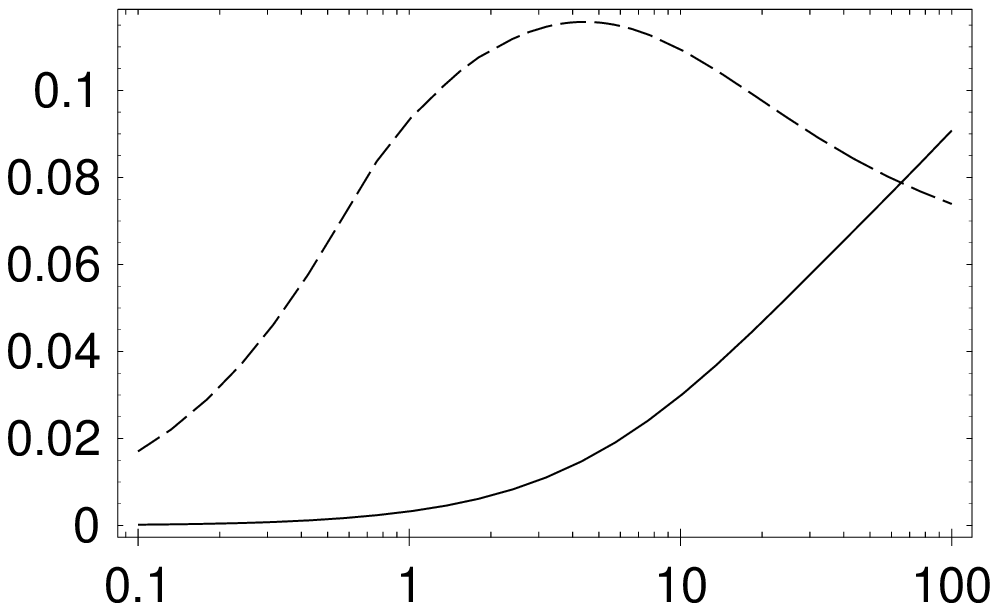}
\begin{picture}(0,0)
\setlength{\unitlength}{1mm}
\put(30,0){\small{$Q^2$ [GeV$^2$]}}
\put(-10,30){$F_2^\ga/\al$}
\end{picture}
\end{minipage}
\begin{minipage}{55mm}
\epsfxsize55mm
\epsffile{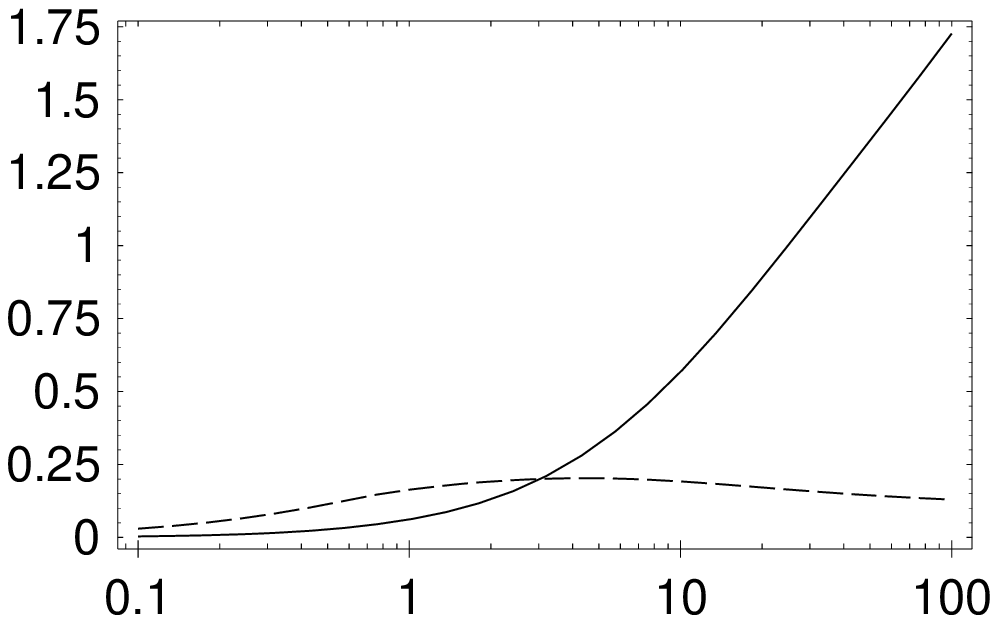}
\begin{picture}(0,0)
\setlength{\unitlength}{1mm}
\put(30,0){\small{$Q^2$ [GeV$^2$]}}
\end{picture}
\end{minipage}
\hspace{-10mm}
\end{center}
\caption{The soft and hard contribution to structure functions at different
values of $x$. Solid line hard contribution from the model; dashed line soft
contribution from the model. First row, proton structure function $F_2$; 
second row, 
longitudinal proton structure function $F_L$ ; third row,  charm
contribution to the proton structure function $F^c_2$: last row, photon
structure function $F^\gamma_2/\alpha$.} \lb{DL}
\end{figure}

With
the rescaling factor (\rf{supp}) and the energy dependence introduced in
(\rf{sidipW}) we can describe the proton structure function, or  equivalently
the total $\gamma^* p$ cross section $\si^{\rm Tot}_{\gamma^* p}$, from $Q^2=0$
up to $Q^2\approx 150$ GeV$^2$. The only free parameter is $R_c$. In figure
\rf{gap} we show some the results for $R_c = 0.22 $ fm compared  with a sample
of experimental data from ZEUS \ct{ZEUS_X} and H1 \ct{H1_X}. We can also 
compare with the fit to the
data in \cite{DL98} where the structure function was  separated into a soft 
and a hard part; the model
reproduces the $Q^2$ dependence of the soft- and hard-pomeron
contributions obtained in \ct{DL98} very well.

With $R_c$ fixed, the model can be used to predict the charm part of the 
proton structure function, $F_2^c(x,Q^2)$, and the longitudinal structure 
function, $F_{L}(x,Q^2)$. In both cases the photon wave function is 
concentrated at smaller distances. In the case of charm this is a
consequence of the mass of the charm quark occuring in
the argument $\ep$ (see (\rf{epsilon})) of the modified
Bessel function in the photon wave functions (\rf{phowa}).
For the longitudinal structure function it is a consequence
of the factor $z(1-z)$ in the wave function of the
longitudinal photon. This factor suppresses contributions from small
values of $\ep$, which correspond to large distances. Thus
the hard pomeron is already dominant at moderate energies, as can be seen from
the second and third rows of figure \rf{DL}. The strong suppression
of the soft pomeron relative to the hard pomeron in $F_2^c(x,Q^2)$,
which is purely a wave-function effect, is notable and provides an
explanation for the almost-complete flavour-blindness of the hard pomeron
commented on in \cite{DL99,DL01}. Comparison of the first and second rows of
figure \rf{DL} shows that for the longitudinal structure function the increase
of the short range (hard part) with increasing $Q^2$ is not as strong as for 
the transverse structure function, a consequence of the less-singular
behaviour of the Bessel functions at the origin in the relevant
photon wave function ($K_0$ {\it vs} $K_1$). Nonetheless as the long range
(soft part) of the longitudinal structure function is even more suppressed
at large $Q^2$ relative to its contribution to the transverse structure 
function, the hard pomeron is dominant sooner in $F_L(x,Q^2)$ than in
$F_2(x,Q^2)$. In figures \rf{f2pch} and \rf{plong} we compare the 
predictions of the model
directly with the experimental results for the charm \cite{ZEUS_XX,H1_XX} 
and longitudinal \cite{H100}
structure functions. The agreement with both data sets is clearly satisfactory.

\begin{figure}
\leavevmode
\begin{center}
\begin{minipage}{4.2cm}
\epsfxsize4.2cm
\epsffile{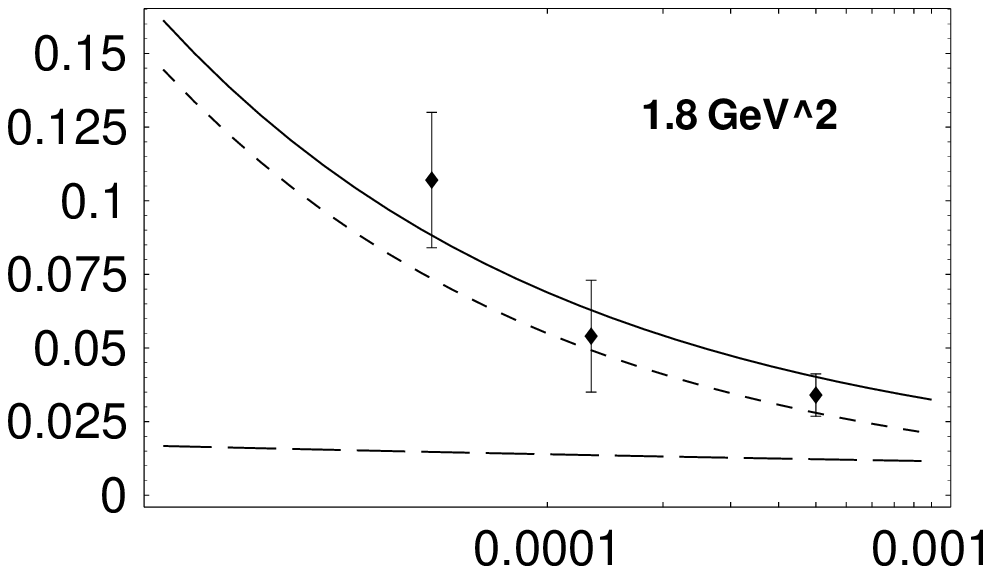}
\end{minipage}
\begin{minipage}{4.2cm}
\epsfxsize4.2cm
\epsffile{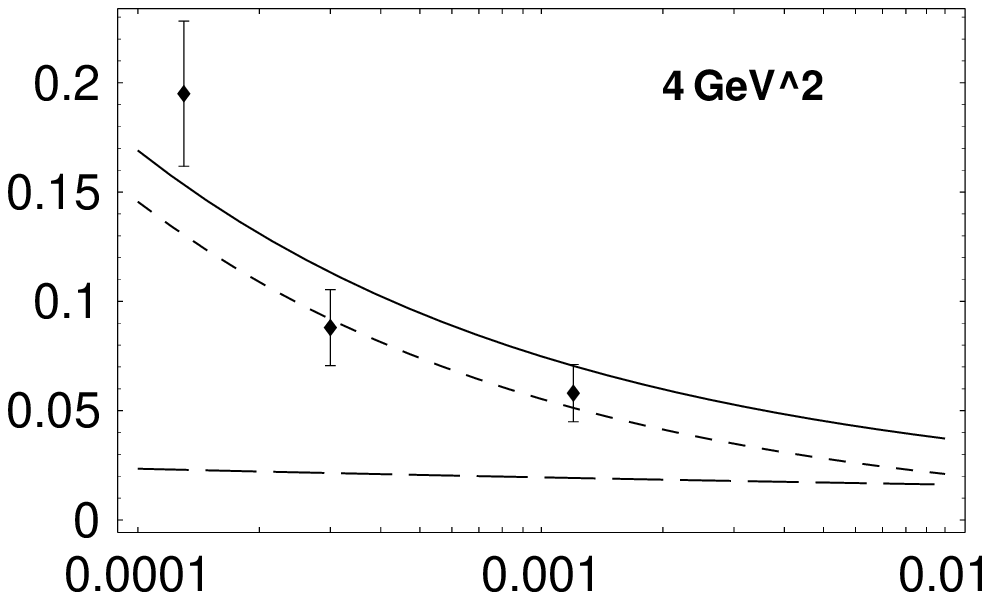}
\end{minipage}
\begin{minipage}{4.2cm}
\epsfxsize4.2cm
\epsffile{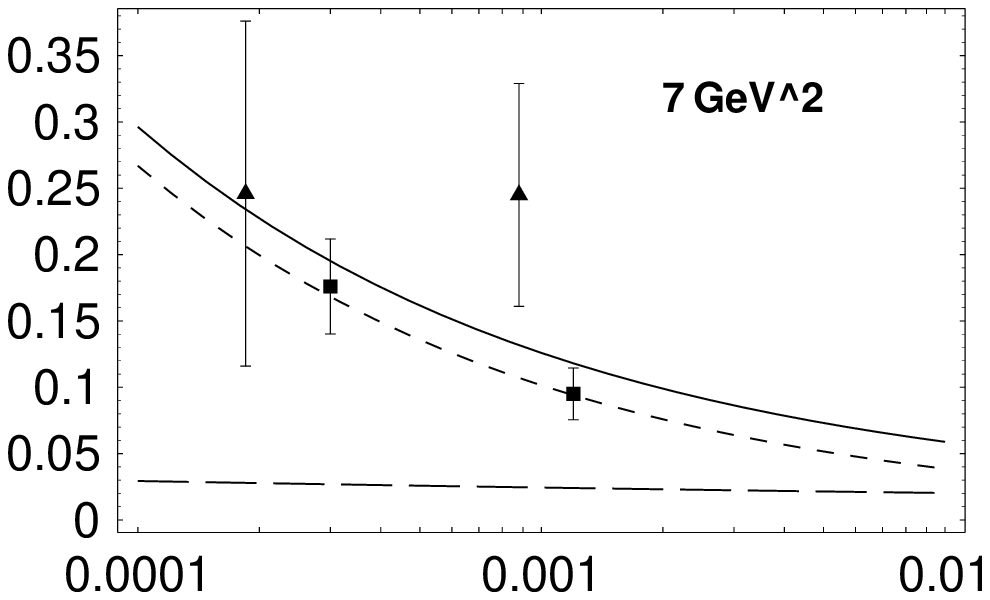}
\end{minipage}
\begin{minipage}{4.2cm}
\epsfxsize4.2cm
\epsffile{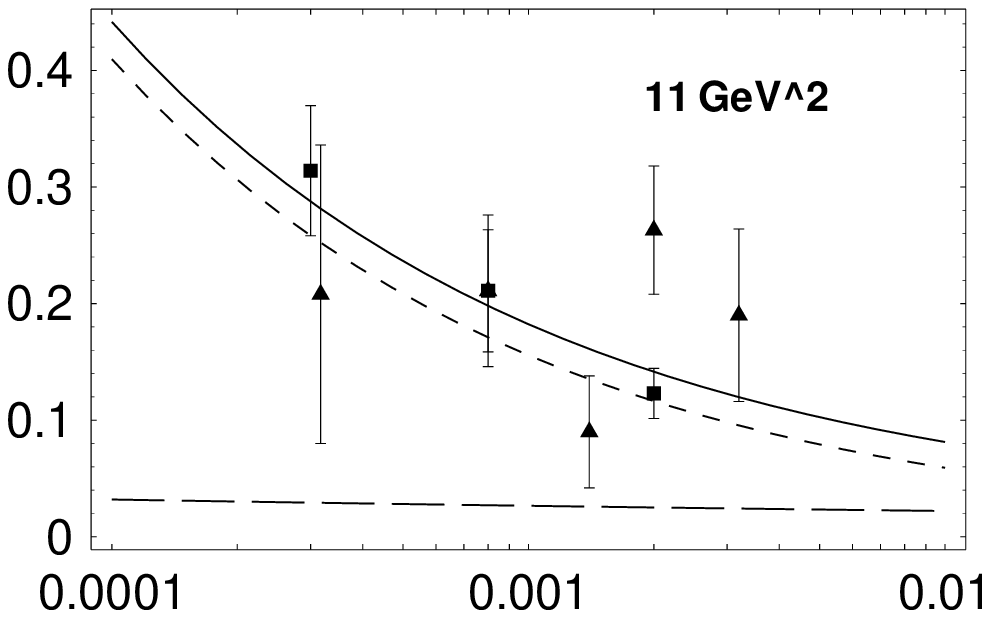}
\end{minipage}
\begin{minipage}{4.2cm}
\epsfxsize4.2cm
\epsffile{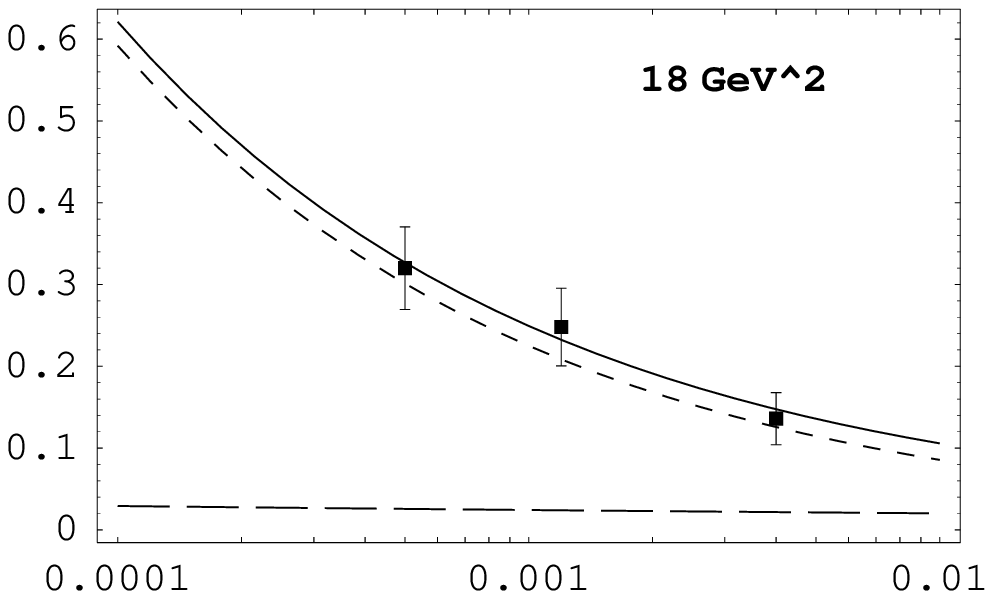}
\end{minipage}
\begin{minipage}{4.2cm}
\epsfxsize4.2cm
\epsffile{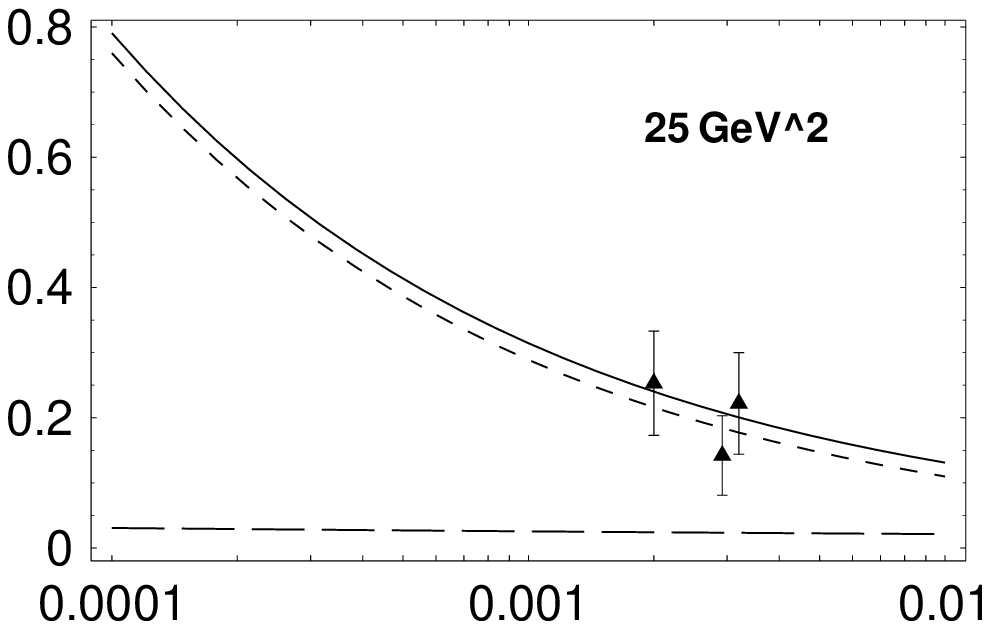}
\end{minipage}
\begin{minipage}{4.2cm}
\epsfxsize4.2cm
\epsffile{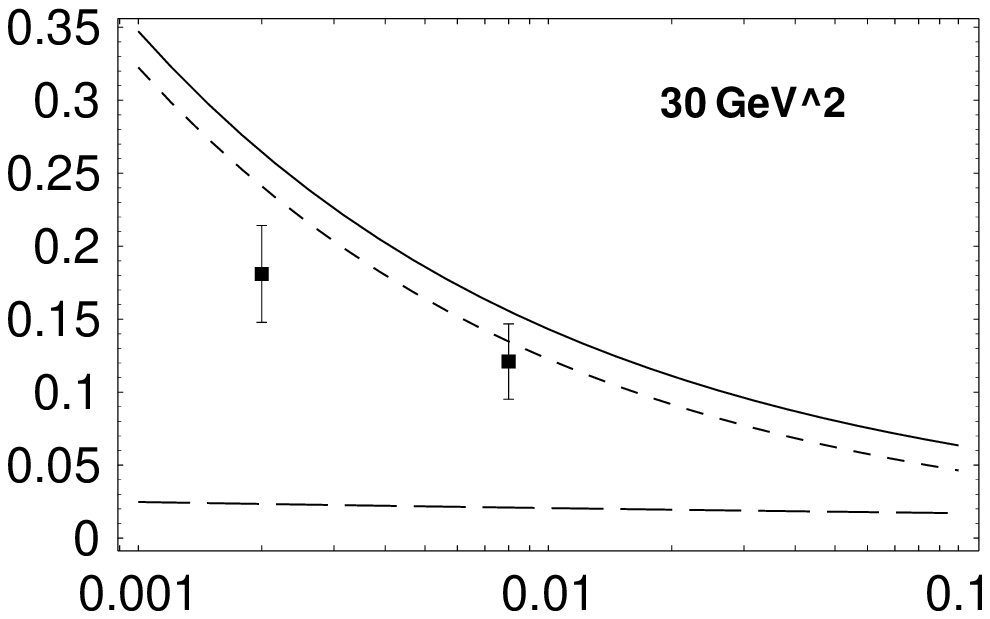}
\end{minipage}
\begin{minipage}{4.2cm}
\epsfxsize4.2cm
\epsffile{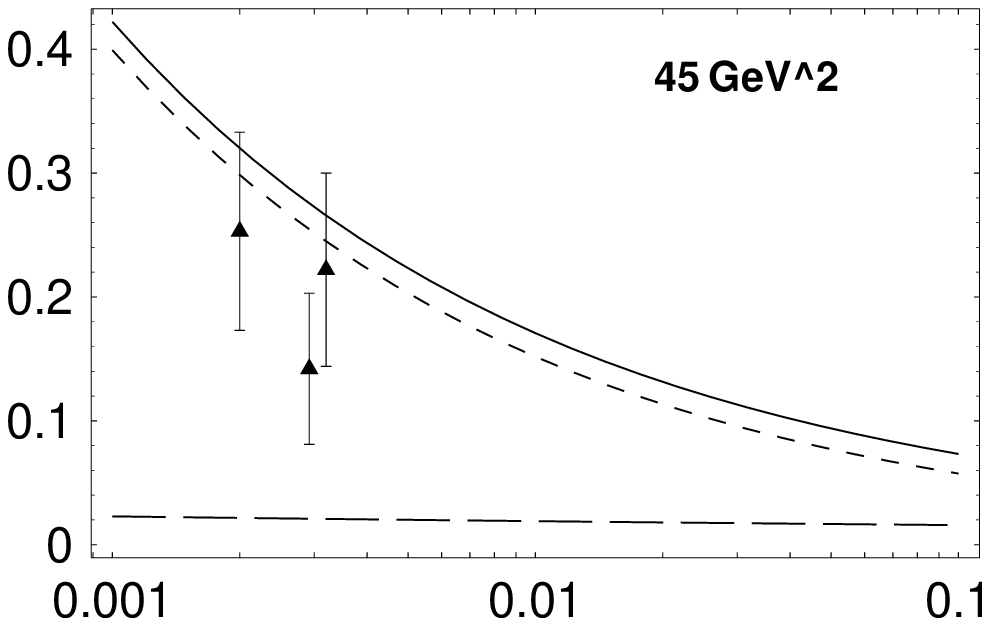}
\end{minipage}
\begin{minipage}{4.2cm}
\epsfxsize4.2cm
\epsffile{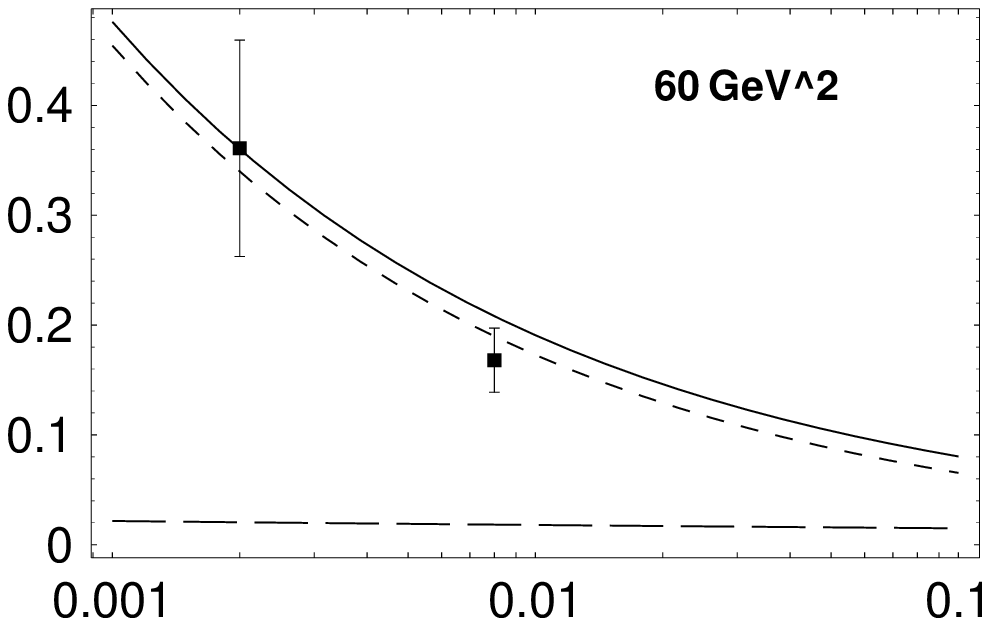}
\end{minipage}
\begin{minipage}{4.2cm}
\epsfxsize4.2cm
\epsffile{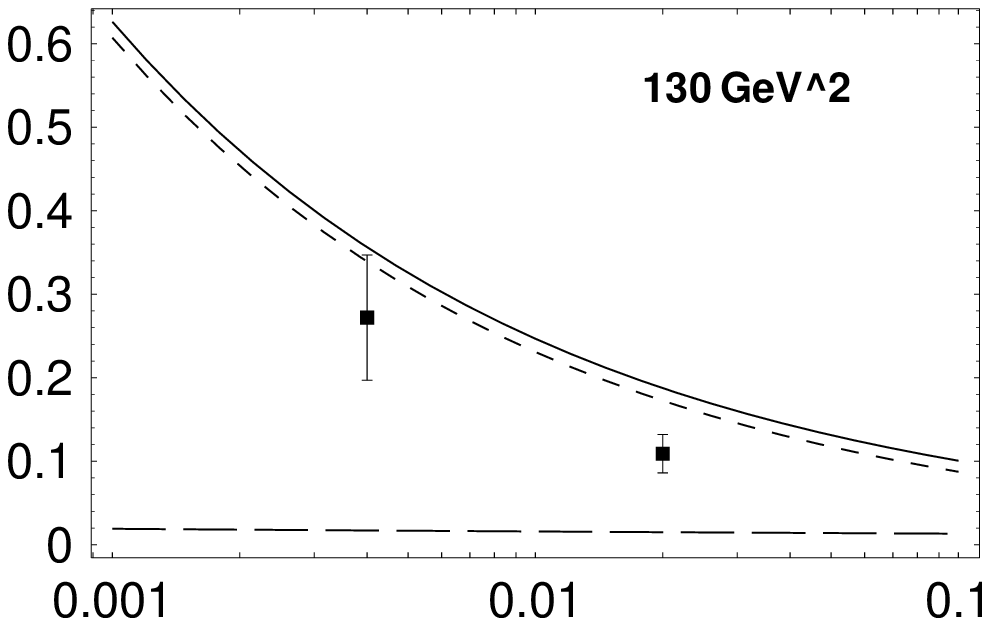}
\begin{picture}(0,0)
\setlength{\unitlength}{1mm}
\put(-7,25){\small $F^c_2$}
\put(30,2){\small $x$}
\end{picture}
\end{minipage}
\end{center}

\caption{The charm contribution to the proton structure function $F^c_2$
for different values of $Q^2$ as indicated in the figures. The solid line 
is the full result, the
short-dashed line the hard pomeron contribution and the long-dashed line
the soft pomeron contribution. The data are from 
ZEUS \protect \ct{ZEUS_XX}, squares, and H1 \protect
\ct{H1_XX}, triangles.}\lb{f2pch} 
\end{figure}
\begin{figure}
\leavevmode
\begin{center}
\begin{minipage}{4.2cm}
\epsfxsize4.2cm
\epsffile{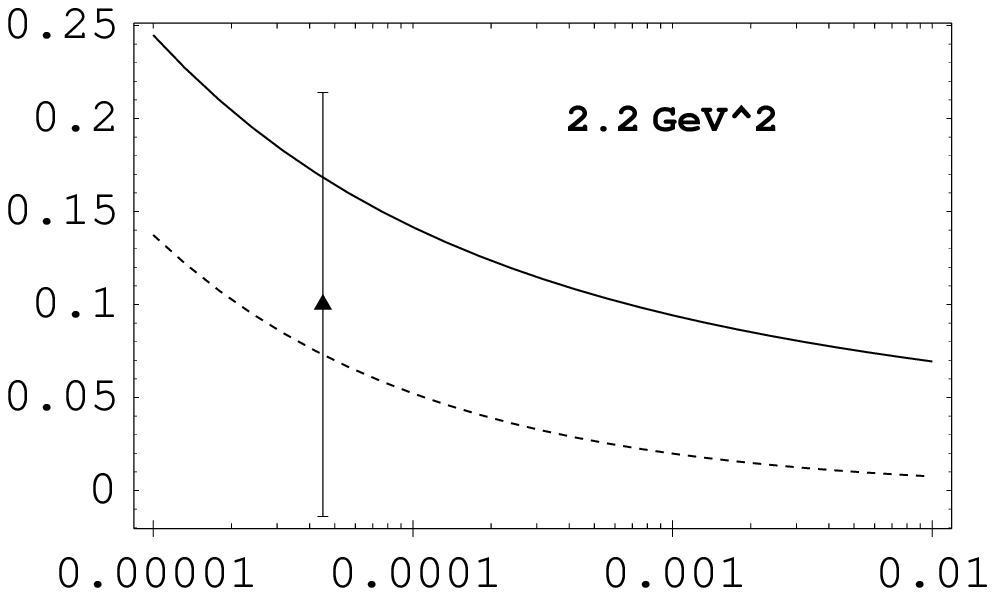}
\end{minipage}
\begin{minipage}{4.2cm}
\epsfxsize4.2cm
\epsffile{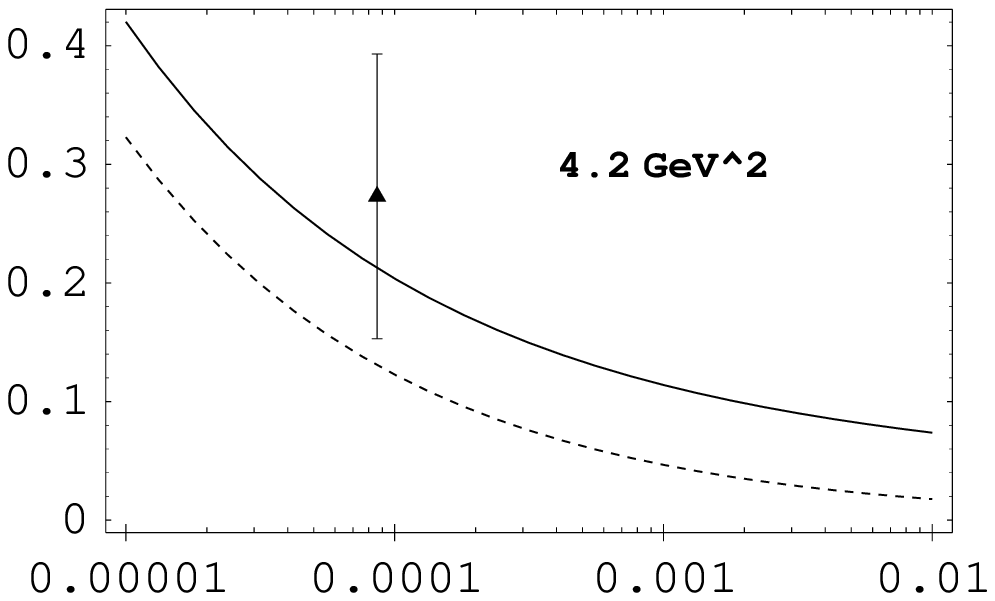}
\end{minipage}
\begin{minipage}{4.2cm}
\epsfxsize4.2cm
\epsffile{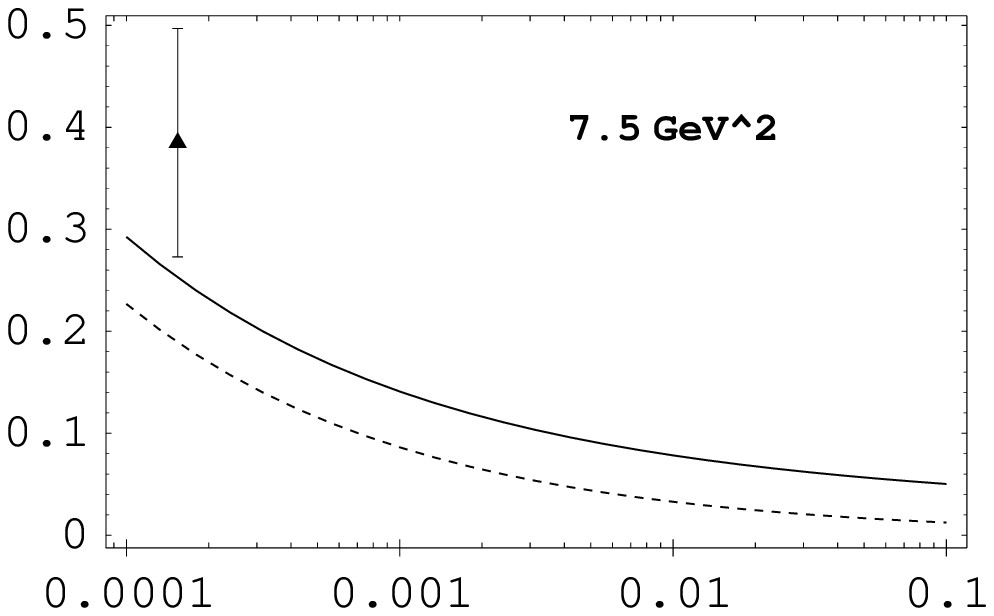}
\end{minipage}
\begin{minipage}{4.2cm}
\epsfxsize4.2cm
\epsffile{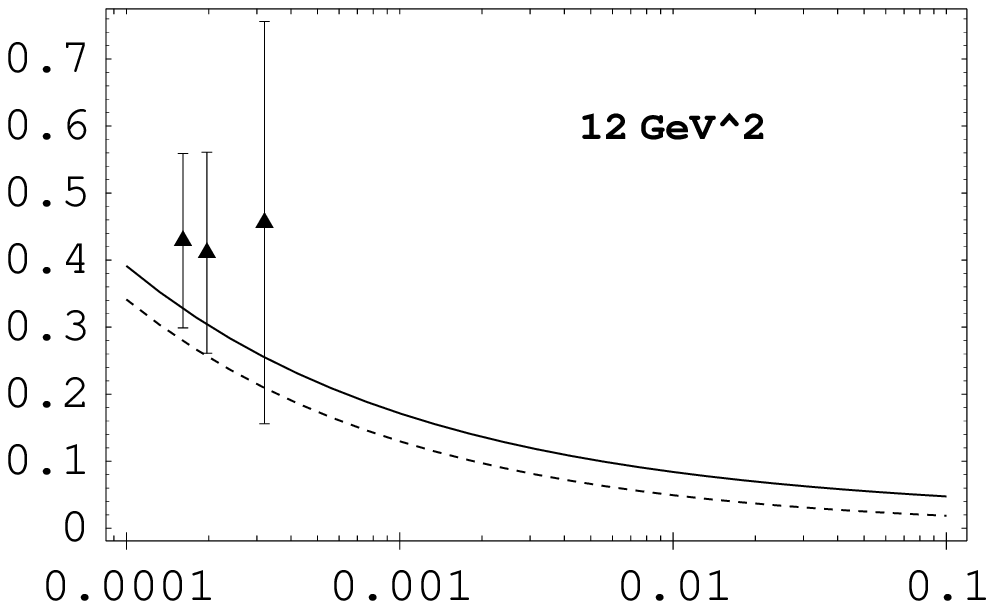}
\end{minipage}
\begin{minipage}{4.2cm}
\epsfxsize4.2cm
\epsffile{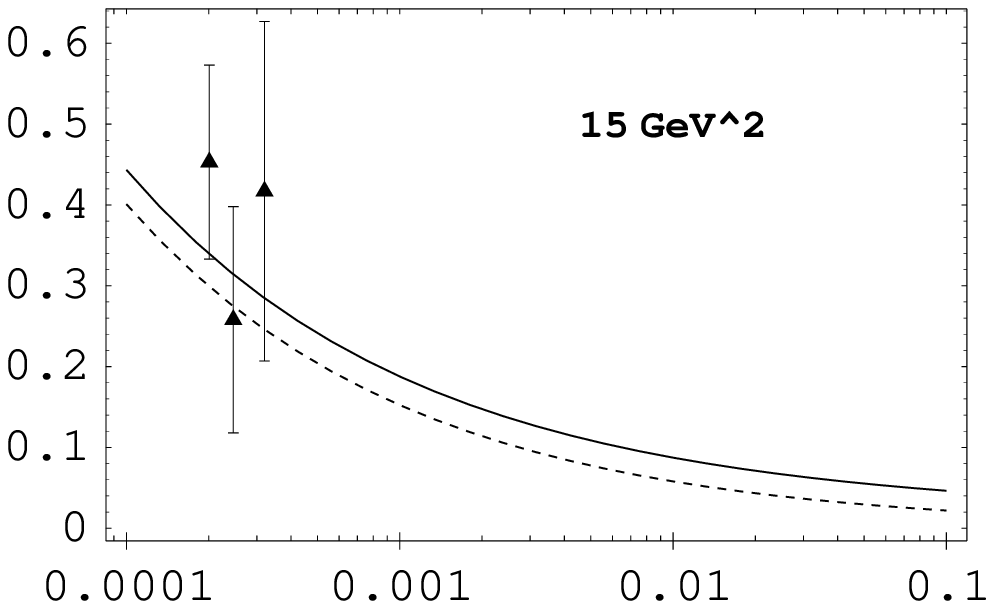}
\end{minipage}
\begin{minipage}{4.2cm}
\epsfxsize4.2cm
\epsffile{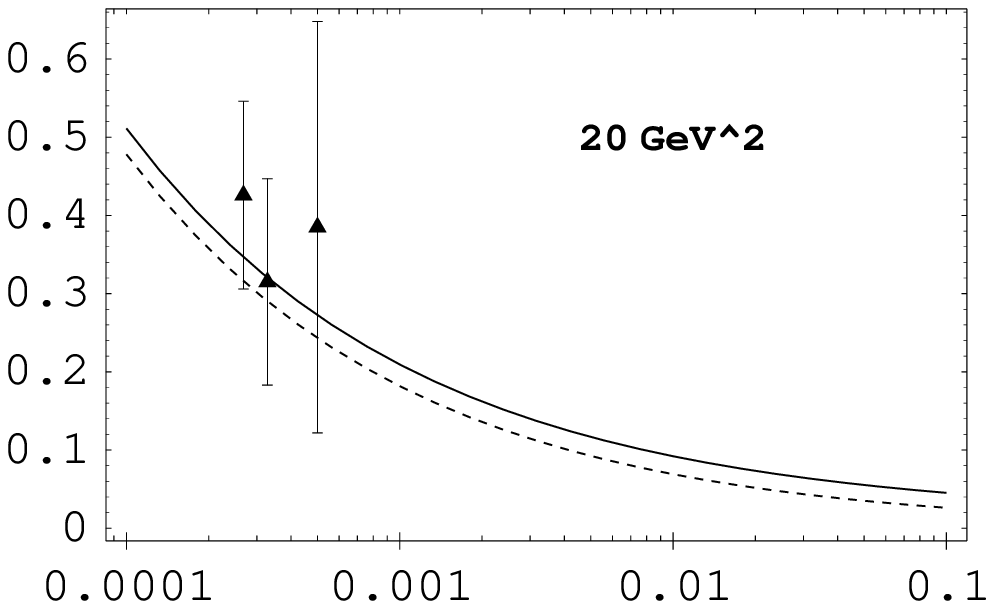}
\end{minipage}
\begin{minipage}{4.2cm}
\epsfxsize4.2cm
\epsffile{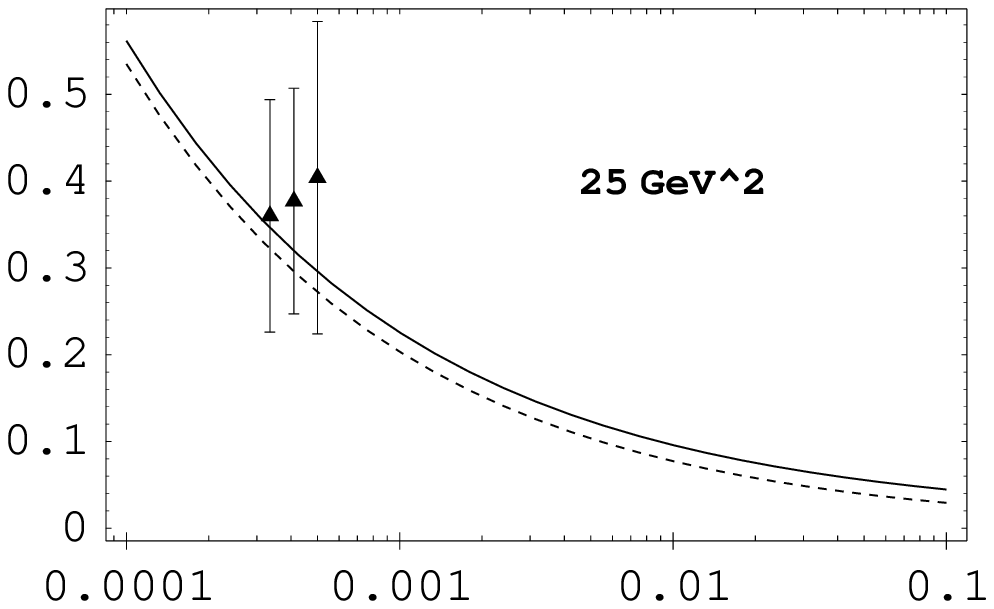}
\begin{picture}(0,0)
\setlength{\unitlength}{1mm}
\put(30,2){\small$x$}
\put(-8,23){\small$F_L$}
\end{picture}
\end{minipage}
\begin{minipage}{4.2cm}
\epsfxsize4.2cm
\epsffile{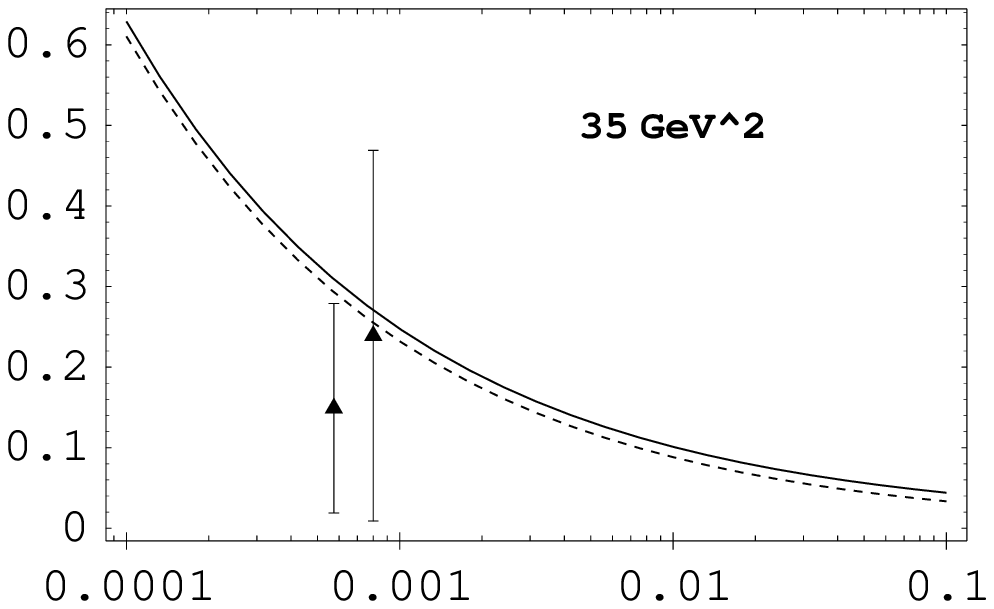}
\begin{picture}(0,0)
\setlength{\unitlength}{1mm}
\put(30,2){\small$x$}
\end{picture}
\end{minipage}
\end{center}
\caption{The longitudinal proton structure function
$F_{L}$ for different values of $Q^2$ as indicated in the
figures. The solid line is the full result and the dashed line
the hard pomeron contribution. The data are from H1 \protect
\ct{H100}.}\lb{plong} 
\end{figure}

It is instructive to compare the photoproduction of $J/\psi$
mesons with the charm structure function $F_2^c$. Experiment tells us 
that the ratio of the soft pomeron to the hard pomeron is much larger for 
$J/\psi$ production than for the charm contribution to the proton structure 
function at comparable energies. This can be
easily understood in the dipole-model approach.
The virtual $c\bar c$ pair in the
photon wave function has an extension of  $\approx 1/m_c$, but the 
$J/\psi$ has a much larger radius, in the range from a typical
hadronic radius to the Bohr radius of order $1/(\al_s m_c)$. 
Therefore the overlap of the charm part of the photon wave function 
with the $J/\psi$ wave function obtains a larger
contribution from distances $R > R_c$ than does the square of the charm
part of the photon wave function. 

In the approach presented here we can only evaluate the forward production
amplitude. In an earlier investigation \cite{DGKP97} with the same model
at a centre-of-mass energy of $W=20$ GeV an effective logarithmic
slope of the production cross section of about 6 GeV$^{-2}$
was found. 
For our
calculation we have used the  same $J/\psi$ wave function as there.
In figure \rf{gapsi} we show the integrated
production cross section for two cases: with a
constant logarithmic slope of $b$=6 GeV$^{-2}$ (dashed line)
and with a slope varying with energy as predicted from Regge theory,
where the trajectory of the soft pomeron has the slope
$\al_{P_s}'=0.25$ GeV$^{-2}$ and that of the hard
pomeron \ct{DL99} is $\al_{P_h}'=0.1$ GeV$^{-2}$ (solid line) with 
$b$= 6 GeV$^{-2}$ at $W=20$ GeV. The agreement with the H1 \ct{H1_J}
and ZEUS \ct{ZEUS_J} data is satisfactory in both cases. The actual
normalization of the cross section is rather sensitive to the special choice 
of the wave function, but the energy dependence is much less so, as can be 
inferred from the general arguments given above.  
\begin{figure} \leavevmode \begin{center}
\begin{minipage}{8cm}
\epsfxsize8cm
\epsffile{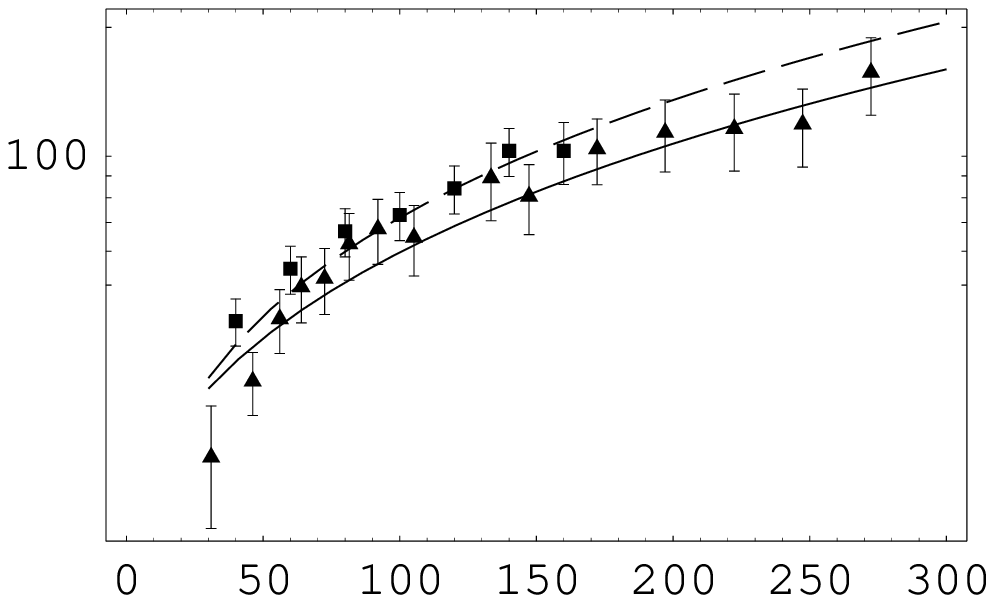}
\begin{picture}(0,0)
\setlength{\unitlength}{1mm}
\put(50,2){$W$ [GeV]}
\put(-5,40){$\si$}
\put(-8,35){[$\mu$b]}
\end{picture}
\end{minipage}
\end{center}
\caption{
Cross section for the reaction $\gamma p \to J/\psi p$.
The long-dashed curve is obtained from the forward amplitude using a constant
logarithmic slope $b = 6$ GeV$^2$ and the solid curve using an $s$-dependent 
slope from Regge theory. The data are from H1 \ct{H1_J},triangles, and 
ZEUS \ct{ZEUS_J}, squares.
} \lb{gapsi}
\end{figure} 

\begin{figure}[t]
\leavevmode
\begin{center}
\hspace{7mm}\begin{minipage}{6cm}
\epsfxsize6cm
\epsffile{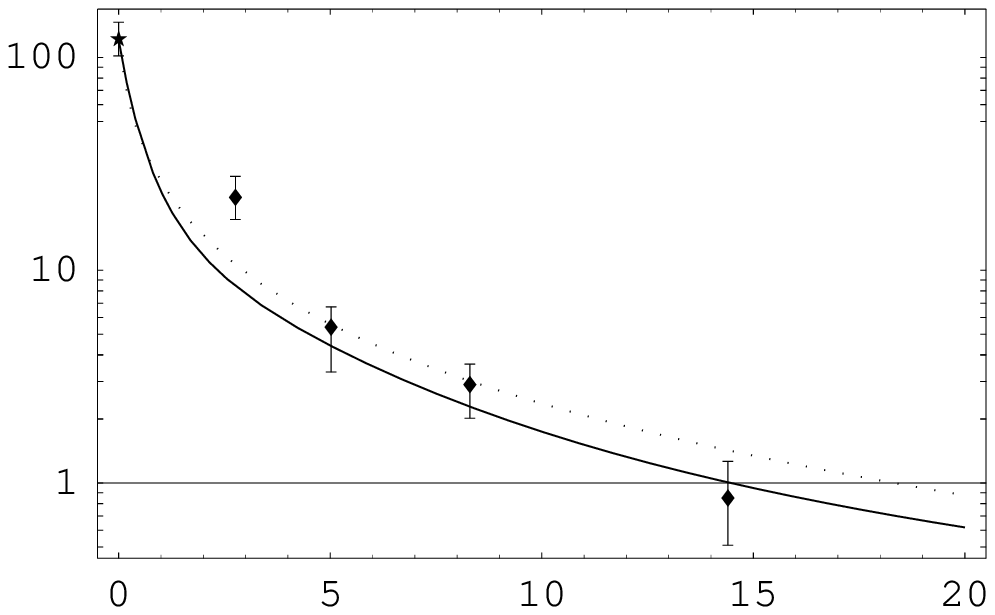}
\begin{picture}(0,0)
\setlength{\unitlength}{1mm}
\put(35,1){$Q^2$ [GeV$^2$]}
\put(-5,30){$\si$}
\put(-8,25){[nb]}
\put(25,-3){(a)}
\end{picture}
\end{minipage}
\hfill
\begin{minipage}{6cm}
\epsfxsize6cm
\epsffile{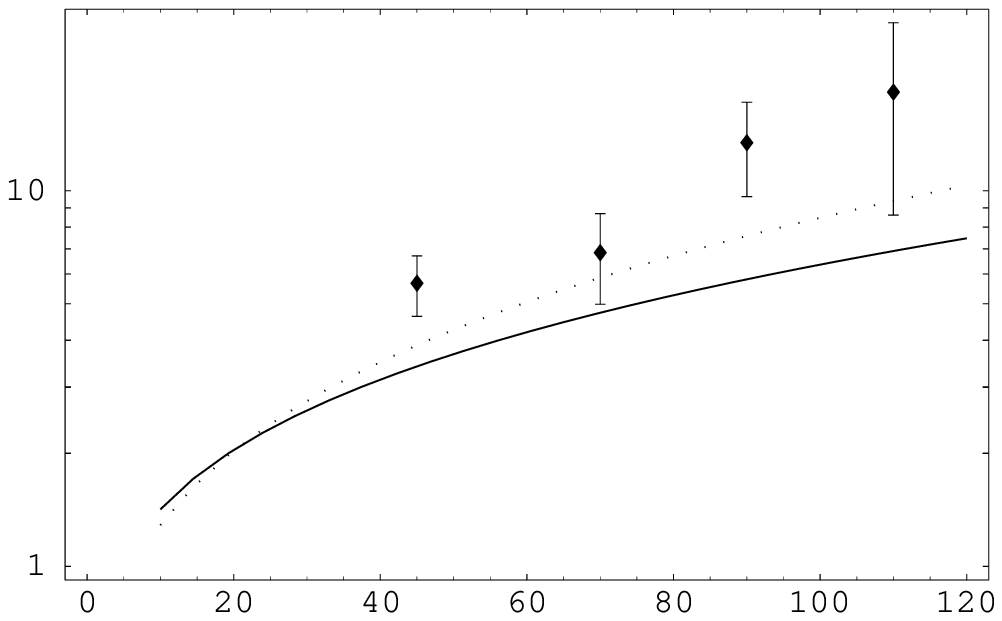}
\begin{picture}(0,0)
\setlength{\unitlength}{1mm}
\put(35,1){$W$ [GeV]}
\put(-5,30){$\si$}
\put(-8,25){[nb]}
\put(25,-3){(b)}
\end{picture}
\end{minipage}
\end{center}
\caption{(a)The integrated cross section for the reaction
$\gamma^*\,p \to \gamma\,p$ as a function of the virtuality
$Q^2$ of the incoming photon at an averaged $\langle W
\rangle= 75$ GeV. (b)The integrated cross section for the
reaction $\gamma^*\,p \to \gamma\,p$ as function of the
centre-of-mass energy W at an averaged virtuality of the
incoming photon of $\langle Q^2\rangle = 4.5$ GeV$^2$. The
solid line is the result of the present form of the model,
the dashed line the result of \protect \ct{DD01} and the data
are preliminary H1 results \ct{H1200}.}
\label{compt} 
\end{figure}

In a recent paper \ct{DD01} we investigated deep virtual Compton
scattering, $\ga^* p\to \ga p$, in essentially the same model as here. 
In that paper, as in 
previous investigations \ct{DDR99,DDR00} on $\ga\ga$ reactions,
we used a somewhat different procedure \ct{Rue99} to incorporate a hard
scale into the nonperturbative model. Instead of the rescaling (\rf{supp}) 
used here, in \ct{DD01} the
dipole cross section (\rf{sidip}) was put to zero if at least one dipole was
smaller than 0.16 fm. In figure \rf{compt} we show  the integrated cross
sections as a function of $Q^2$ and of $W$. The solid  line is our present
prediction compared with the result for the previous procedure (dotted line).
The data are the preliminary H1 data \ct{H1200} after subtraction of the
Bethe-Heitler contribution.
Also in this case only the forward scattering  amplitude has been calculated.
For comparison with experiment the  integrated cross section has been obtained
assuming a constant logarithmic slope $b=7$ GeV$^{-2}$ which is the average
value over the $Q^2$ range of the preliminary data. The only serious 
discrepancy between the model and the preliminary data is at $Q^2 = 3.5$ 
GeV$^2$. This is reflected in the normalisation of the integrated cross 
section in figure \ref{compt}(b) as the low-$Q^2$ point dominates this cross 
section. Note, however, the important constraint put on models by the real
$\gamma p$ data. The model prediction is compared with the real $\gamma p$ cross
section $\sigma^{\rm Tot}_{\ga p}$ in figure \rf{sipgare}. Note that the model 
predicts a significant contribution from the hard pomeron to $\sigma^{\rm
Tot}_{\ga p}$, similar to that found in \ct{DL98}. However the data do not
demand such a contribution as, due to the comparatively large errors at high
energy, the data can accomodate the standard soft-pomeron energy dependence.
 
\begin{figure}
\leavevmode
\begin{center}
\begin{minipage}{8cm}
\epsfxsize8cm
\epsffile{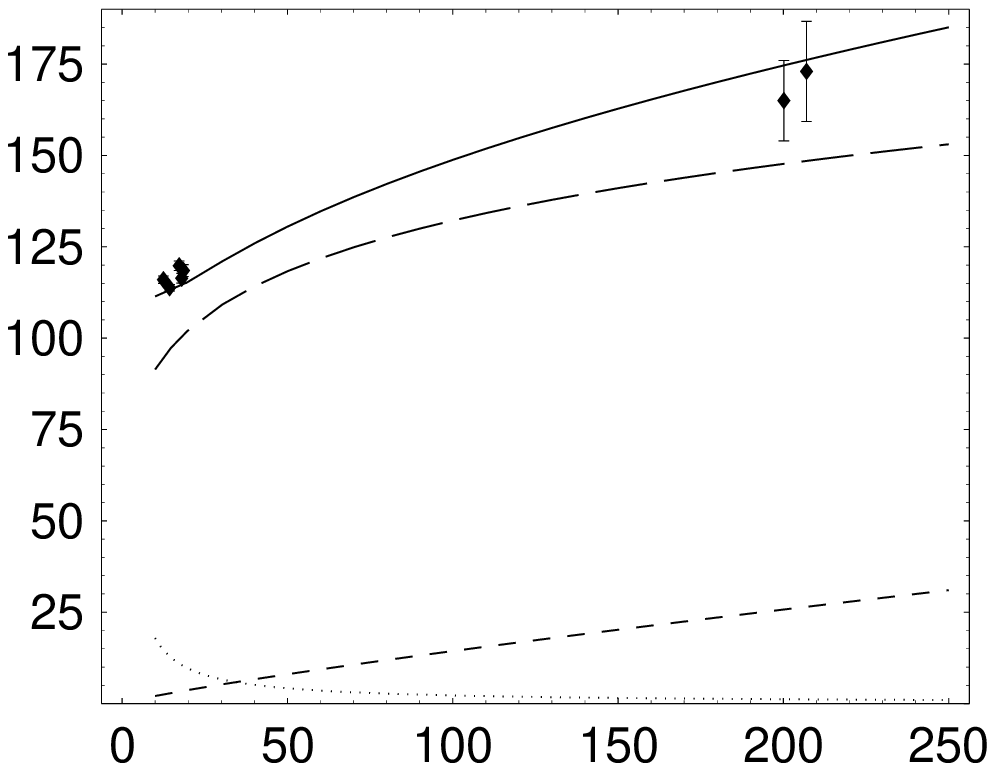}
\begin{picture}(0,0)
\setlength{\unitlength}{1mm}
\put(50,2){$W$ [GeV]}
\put(-5,50){$\si$}
\put(-8,45){[$\mu$b]}
\end{picture}
\end{minipage}
\end{center}
\caption{$\si^{\rm Tot}_{\gamma p}$. The solid line is the full
result. It has the following contributions: long dashes, soft pomeron; short
dashes, hard pomeron; dots, reggeon.
The data are from \protect \ct{gamma-p}} \lb{sipgare} \end{figure}

\clearpage

\subsection{$\gamma$-$\gamma$ reactions}

\begin{figure}
\leavevmode
\begin{center}
\begin{minipage}{120mm}
\epsfxsize120mm
\epsffile{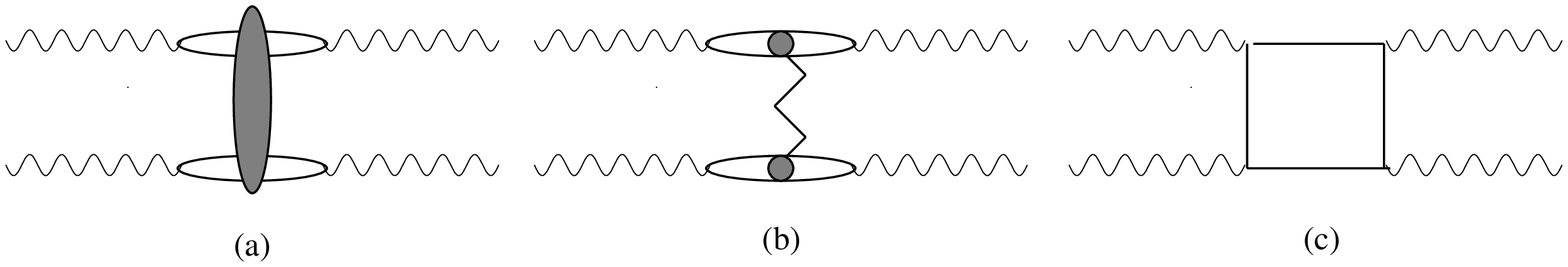}
\end{minipage}
\end{center}
\caption{Graphical representation of the dipole-dipole model contribution (a),
the reggeon contribution (b) and the box diagram (fixed pole, quark parton
model), (c)} \label{gaga}  \end{figure}
\begin{figure}
\leavevmode
\begin{center}
\begin{minipage}{8cm}
\epsfxsize8cm
\epsffile{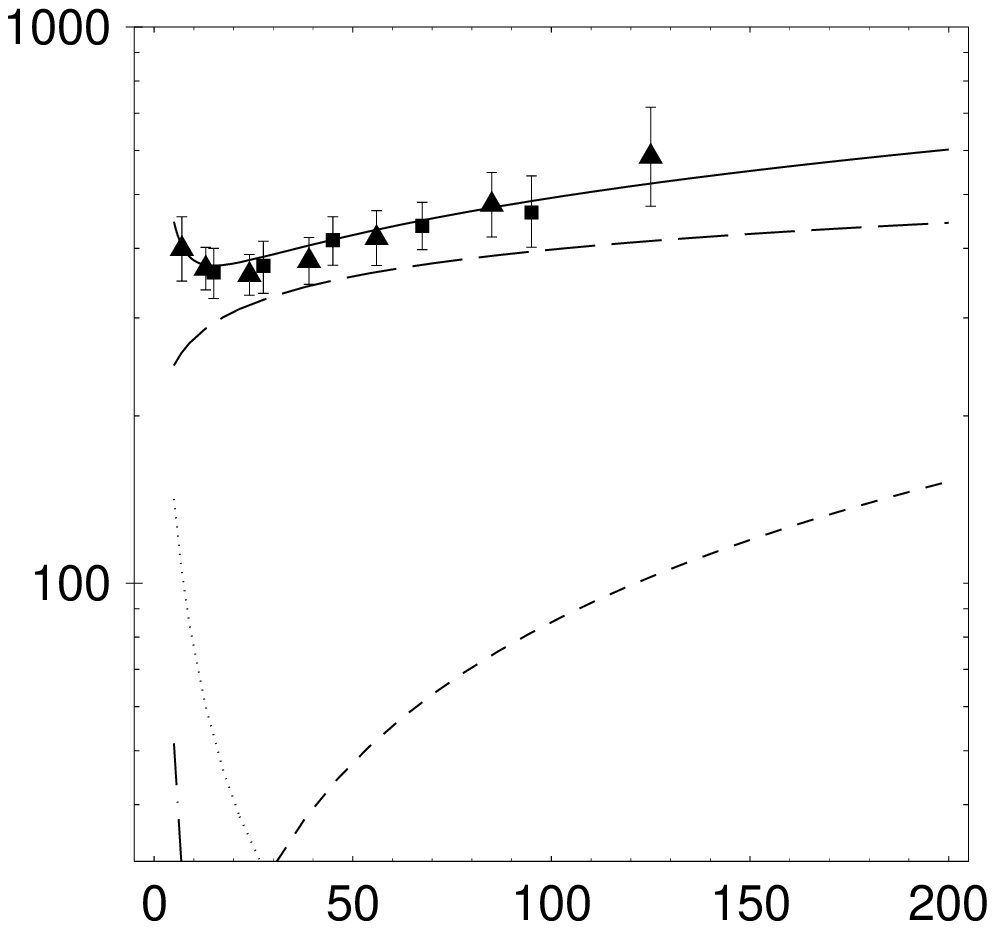}
\begin{picture}(0,0)
\setlength{\unitlength}{1mm}
\put(50,0){$W$ [GeV]}
\put(0,70){$\si$}
\put(-3,65){[$\mu$b]}
\end{picture}
\end{minipage}
\end{center}
\caption{
$\sigma^{\rm Tot}_{\gamma\gamma}(W)$. The solid line
is the full result. The separate contributions are: long dashes, soft pomeron;
short dashes, hard pomeron; dot-dashes, fixed pole (box diagram); dots,
reggeon. The data are from  OPAL\protect \ct{OPAL00a}, boxes, and L3 \protect
\ct{L301}, triangles.} \lb{sigare}
\end{figure} 

With the same approach and the same parameters we can also
calculate $\ga\ga$, $\ga^*\ga$ and $\ga^*\ga^*$ cross sections. 
Since some of the experimental results are
obtained at relatively low centre-of-mass energies the reggeon
contribution, figure \rf{gaga}(b), and the box diagram, figure \rf{gaga}(c),
have to be taken into account. As an estimate for the reggeon contribution we
use the form given in \cite{DDR00}. It was pointed out there that there is
considerable uncertainty in this contribution. We include this uncertainty 
when comparing our predictions with data.
Analytical results for the box diagram 
without any approximations can be found in \cite{BGMS75}. In the framework of
Regge theory it correponds to a fixed pole in the angular momentum plane and
has therefore to be added to the moving Regge-pole contribution \cite{DDR00}. In
the literature it is  often quoted as quark-parton-model (QPM) contribution. 

The principal difference between $\gamma \gamma$ and $\gamma p$
at high energies comes from  the singularity of
the photon wave function at the origin. This favours
the hard component and therefore it should become
apparent even in the scattering of real photons. In figure
\rf{sigare} we show our result for the cross section $\si^{\rm Tot}_{\ga\ga}$
together with OPAL\ct{OPAL00a} and L3\ct{L301} data.
The experimental cross sections are rather sensitive to the Monte Carlo 
model used for the unfolding of detector effects, different Monte Carlos 
producing different results. The resulting uncertainty is contained in 
the errors on the OPAL data. The L3 data shown are the average of the two
extremes. In this case the energy dependence of the data is not compatible
with the soft pomeron alone, and the additional contribution of the hard
pomeron is required.

\begin{figure}
\leavevmode
\begin{center}
\begin{minipage}{60mm}
\epsfxsize60mm
\epsffile{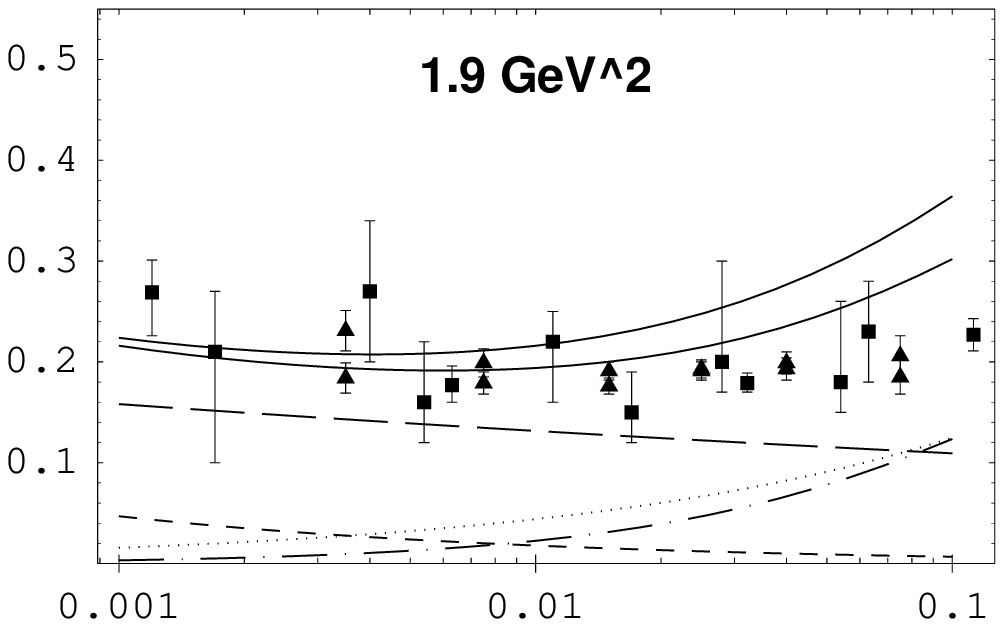}
\begin{picture}(0,0)
\setlength{\unitlength}{1mm}
\put(-7,30){\small $F^\ga_2$}
\end{picture}
\end{minipage}
\begin{minipage}{60mm}
\epsfxsize60mm
\epsffile{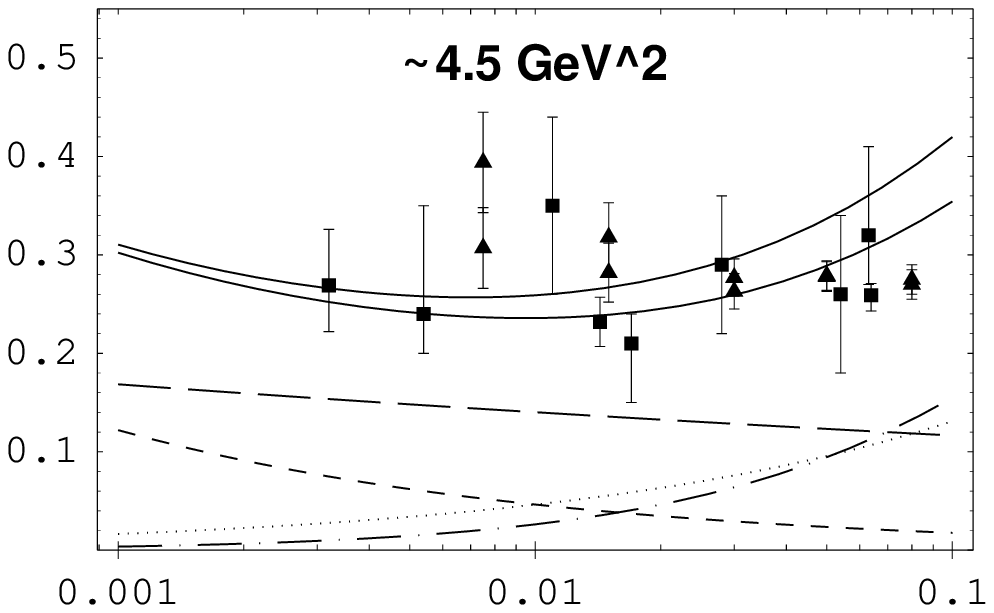}
\begin{picture}(0,0)
\setlength{\unitlength}{1mm}
\end{picture}
\end{minipage}
\begin{minipage}{60mm}
\epsfxsize60mm
\epsffile{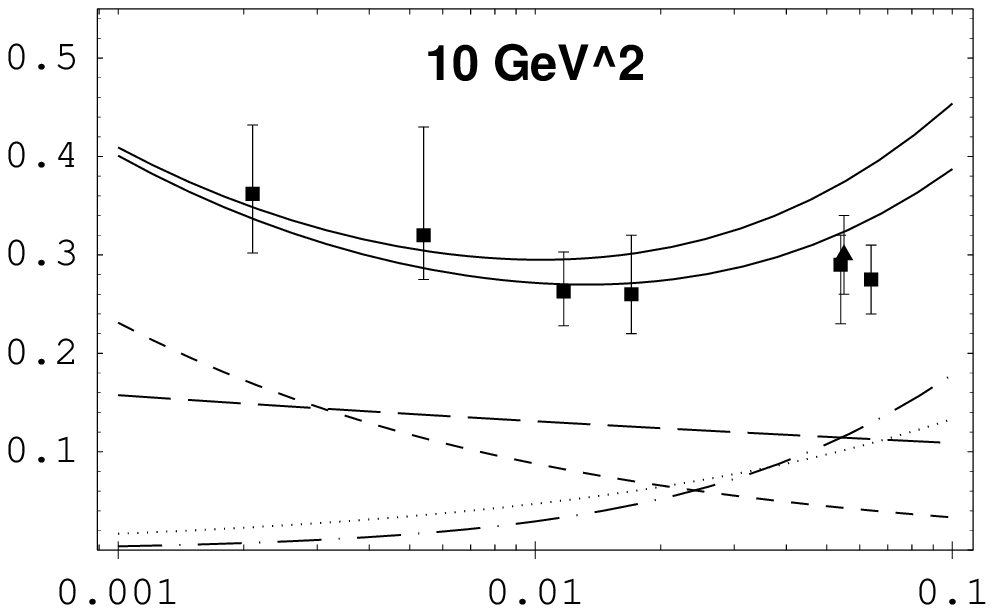}
\begin{picture}(0,0)
\setlength{\unitlength}{1mm}
\put(50,2){\small $x$}
\put(-7,30){\small $F^\ga_2$}
\end{picture}
\end{minipage}
\begin{minipage}{60mm}
\epsfxsize60mm
\epsffile{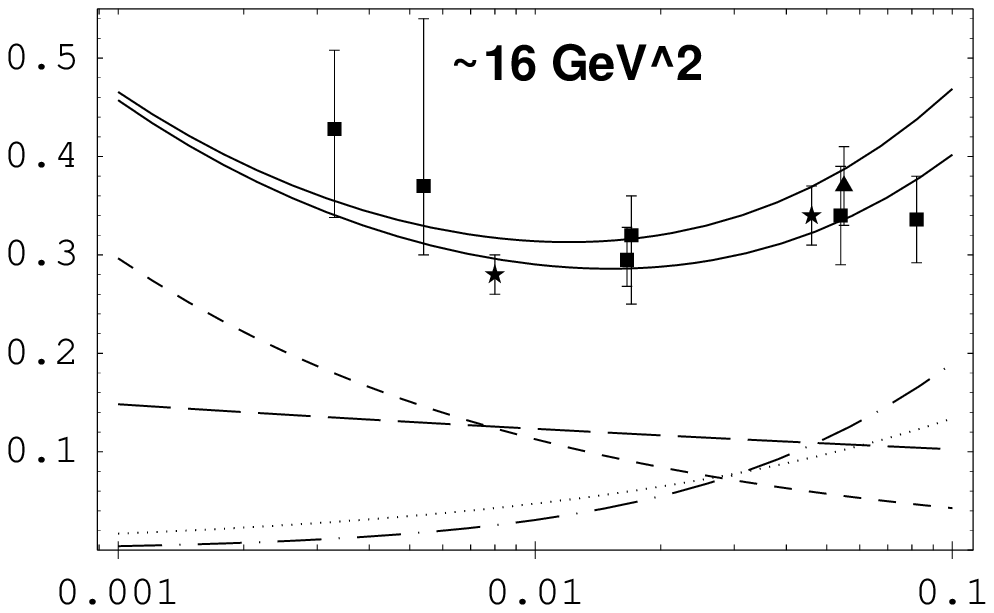}
\begin{picture}(0,0)
\setlength{\unitlength}{1mm}
\put(50,2){\small $x$}
\end{picture}
\end{minipage}
\end{center}
\caption{The photon structure function $F^\gamma_2/\al$
for different values of $Q^2$ as indicated in the figures. 
The upper solid line includes the ``full Regge'', the  lower solid
line includes ``half Regge''. The separate contributions are: long dashes, 
soft pomeron; short dashes, hard pomeron; dot-dashes,  fixed pole (box 
diagram); dots, reggeon. The data are from OPAL \protect 
\ct{OPAL00b,OPAL97a,OPAL97b},
boxes; L3 \protect \ct{L398,L399}, triangles; and 
ALEPH \protect  \ct{ALEPH99}, stars.}\lb{fga} 
\end{figure} 

The model predictions for $F^\ga_2/\alpha$ are equally satisfactory.
A comparison with data is made in figure \ref{fga}.  The agreement with 
experiment is good for small values of $x$. At large $x$ the increasing
importance of the Regge term induces an increasing uncertainty in the
predictions, but nonetheless they remain satisfactory for $x \le 0.1$.
The model predictions for the shape of the photon structure function
$F^\ga_2/\alpha$ are very similar to those for the proton structure function, 
as can be seen by comparing the first and last rows of figure \rf{DL}. 
Indeed at large $Q^2$ the
photon structure function exhibits precisely the same sensitivity to the hard
contribution as does the proton structure function. To quantify this,
in figure \rf{dlpga} we display the ratio $R = (F^\ga_2/\al)/F_2$
for the soft and the hard contributions separately as a function of $Q^2$.
The ratio of the soft contributions (dashed line) is practically constant.
The hard contribution to $F^\ga_2$ is relatively favoured at small $Q^2$, 
a consequence of the singularity of the photon wave function at the
origin, but the ratio of the hard contributions 
tends to the same constant as that of the soft contributions at large
$Q^2$. Of course this is a 
model-dependent statement, but nonetheless it emphasises the importance of
$\sigma^{\rm Tot}_{\ga\ga}$ as a probe of the hard contribution.

\begin{figure}
\leavevmode
\begin{center}
\begin{minipage}{80mm}
\epsfxsize80mm
\epsffile{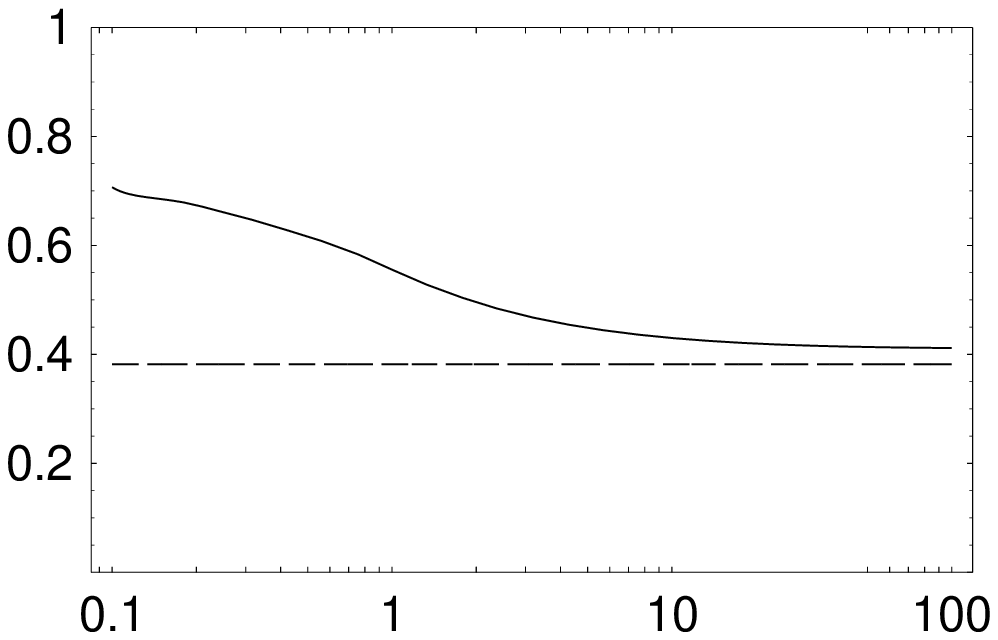}
\begin{picture}(0,0)
\setlength{\unitlength}{1mm}
\put(50,2){\small$Q^2$~[GeV$^2$]}
\put(-7,50){\small$R$}
\end{picture}
\end{minipage}
\end{center}
\caption{$R$, the ratio of the soft and hard contributions of the 
photon  to
the proton structure function. The solid line is the ratio of the hard 
contributions, the dashed line the ratio of the soft.} \label{dlpga} 
\end{figure}

\begin{figure}
\leavevmode
\begin{center}
\begin{minipage}{8cm}
\epsfxsize8cm
\epsffile{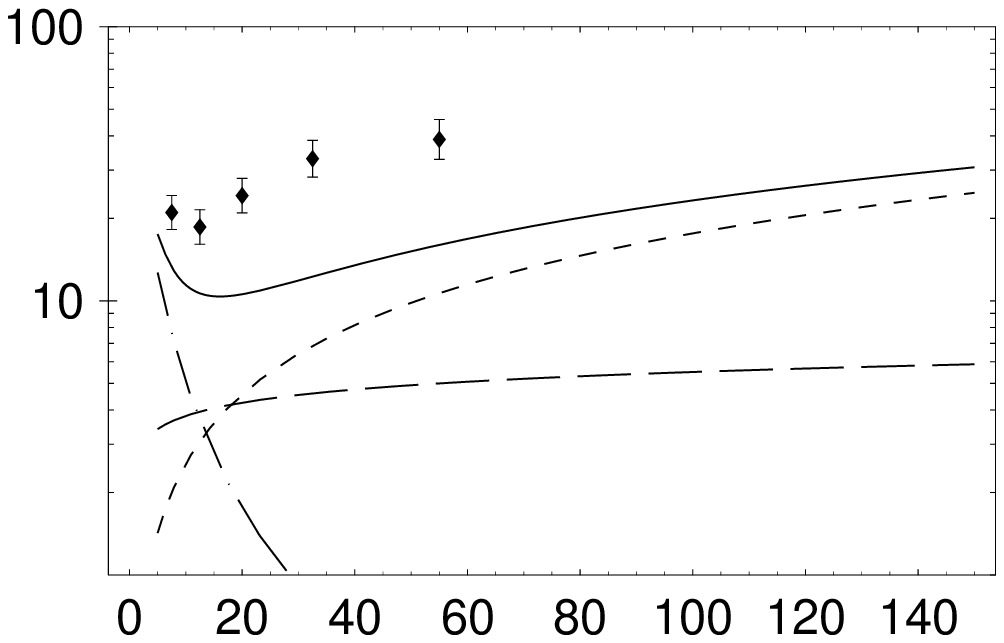}
\begin{picture}(0,0)
\setlength{\unitlength}{1mm}
\put(50,2){$W$ [GeV]}
\put(-8,40){$\si$}
\put(-12,35){[$\mu$b]}
\end{picture}
\end{minipage}
\end{center}
\caption{ Cross section for the reaction
$\gamma \,\gamma \to c \bar c X $. The solid line is the full result.
The separate contributions are: long dashes, soft
pomeron; short dashes, hard pomeron; dot-dashes,  fixed pole (box
diagram). The data are from 
L3 \protect \ct{L300a}.}\lb{sigarech}
\end{figure}

\begin{figure}
\leavevmode
\begin{center}
\begin{minipage}{8cm}
\epsfxsize8cm
\epsffile{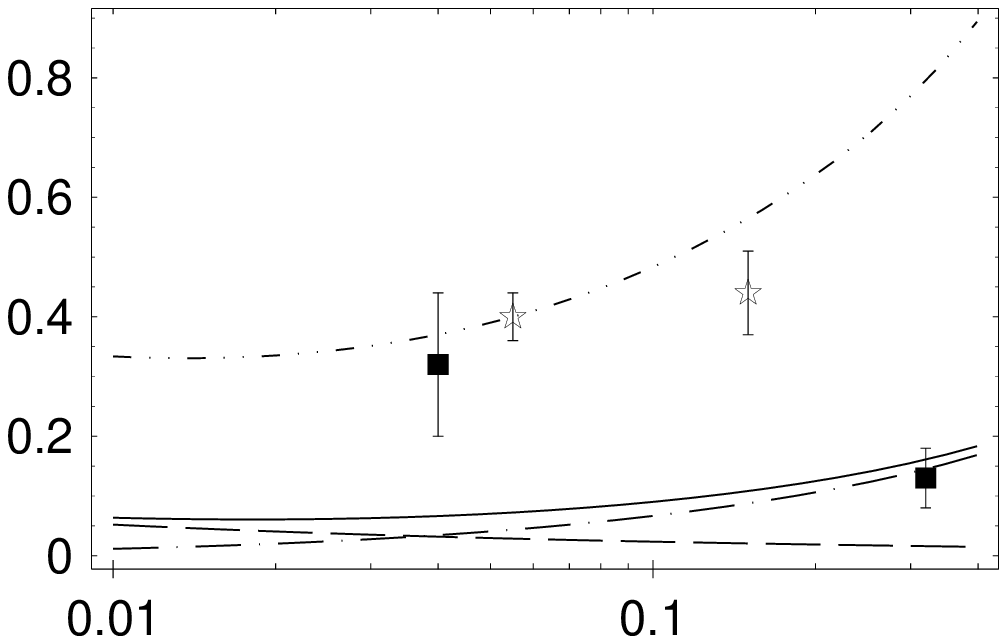}
\begin{picture}(0,0)
\setlength{\unitlength}{1mm}
\put(50,2){$x$}
\put(-10,40){$F^\ga_2/\al$}
\end{picture}
\end{minipage}
\end{center}
\caption{Charm contribution to the photon structure function
$F^\gamma_{2\,c}/\alpha$. The solid line is our prediction at $Q^2 = 20$
GeV$^2$, the dashed line the hard pomeron contribution and the dotted line
the contribution from the box diagram. The squares are the OPAL data
\protect \ct{OPAL99a} for  charm at $\langle Q^2 \rangle = 20$ GeV$^2$. The
stars are the L3 results for the full photon structure function $F^\gamma_2$ at
$\langle Q^2 \rangle = 23$  GeV$^2$, and the upper curve our model result for
the full photon structure  function $F^\gamma_2/\alpha$ at 23 GeV$^2$.}
\lb{fgach} \end{figure}

There is an interesting discrepancy between the model and the 
experimental results for charm production in $\gamma\gamma$ interactions. 
Whereas we found good agreement with experiment for the charm structure
function of the proton, see figure \rf{f2pch}, our model predictions
for the reaction $\gamma\gamma \to c\bar c X$ are about a factor of 
2 lower than the L3 results \ct{L300a}, as can be seen from figure 
\rf{sigarech}. There is a similar discrepancy with the OPAL \ct{OPAL99a} 
result for the charm contribution to the photon structure function at
small $x$, see figure \rf{fgach}, but not at larger $x$ where the 
box diagram dominates. The small-$x$ datum, taken at face-value,
implies that the charm contribution is already at, or close to, its
asymptotic fraction of the photon structure function. Indeed, within the
erors, it exhausts the full structure function.

\begin{figure}
\leavevmode
\begin{center}
\hspace{10mm}\begin{minipage}{60mm}
\epsfxsize60mm
\epsffile{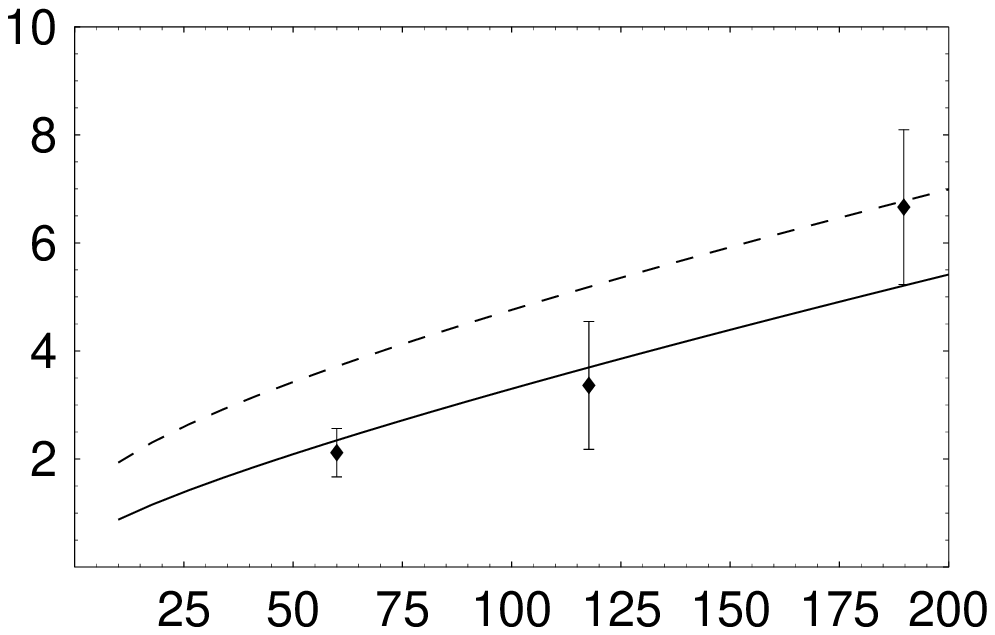}
\begin{picture}(0,0)
\setlength{\unitlength}{1mm}
\put(35,1){$W$ [GeV]}
\put(-5,40){$\si$}
\put(-8,35){[nb]}
\put(25,-3){(a)}
\end{picture}
\end{minipage}
\hfill
\begin{minipage}{60mm}
\epsfxsize60mm
\epsffile{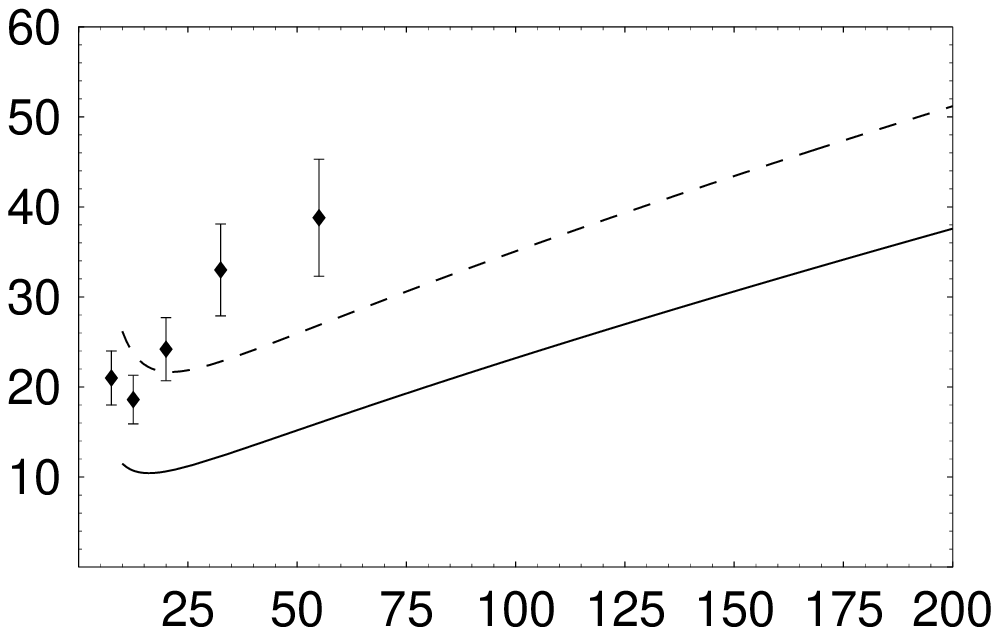}
\begin{picture}(0,0)
\setlength{\unitlength}{1mm}
\put(35,1){$W$ [GeV]}
\put(-5,40){$\si$}
\put(-8,35){[nb]}
\put(25,-3){(b)}
\end{picture}
\end{minipage}
\end{center}
\caption{Model prediction and upper estimates for the production of
charm in (a) $\ga^* p$ and (b) $\ga \ga$ reactions. 
Solid line the actual model prediction, dashed line the upper estimate from
flavour charge independence.The data are from  \protect \ct{ZEUS_XX} (a), and 
L3 \protect \ct{L300a}(b)}
\lb{chnch}
\end{figure}

The discrepancy cannot be removed by simple adjustments of the parameters 
in the model, which are anyway rather tightly constrained by other data.
This can be seen from the following model-independent considerations. At 
very high $Q^2$ charm production in $\gamma^*\gamma$ scattering should be 
$\textstyle{\frac{4}{5}}$ of the total cross section. A factor of 
$\frac{2}{5}$ comes from the ratio of
the square of the charm charge to the sum of the squares of the charges of
all contributing flavours, and there is an additional factor of 2 since the 
$ c{\bar c}$ pair can be created by either photon.
For moderate $Q^2$ one
has to take account of the charm mass and make the replacement $Q^2 \to
Q^2_{\rm eff}$. The latter is the average of the expression 
$Q^2+m_c^2/(z(1-z))$ occuring in the overlap integrals.
The cross section $\textstyle{\frac{4}{5}}\si^{\rm Tot}_{\ga^*\ga}(Q_{eff}^2)$
is then an estimate for charm production in $\ga\ga$ interactions. 
A similar argument can be
applied to $\ga p$ reactions, where $\textstyle{\frac{2}{5}}\si^{\rm
Tot}_{\ga^* p}(Q^2_{eff})$ is the corresponding estimate for charm
production with a photon of virtuality $Q^2$.  As the product $z(1-z)\leq
\textstyle \frac{1}{4}$ then $Q^2+4 m_c^2$ is the lower bound of $Q^2_{eff}$
and therefore the cross sections with that virtuality provide an upper 
estimate for
charm production. In figure \rf{chnch}(a) we show the model prediction for the
cross section for charm production off protons at $Q^2 =1.8$ GeV$^2$ and
compare it with the upper estimate $\textstyle{\frac{2}{5}}\si^{\rm
Tot}_{\ga^* p} (Q^2+4 m_c^2)$ and the experimental data from ZEUS
\cite{ZEUS_XX}. The comparison of the model with the estimate is reasonable
and we note that the estimate indeed tends 
to be above the data. In figure \rf{chnch}(b), where the target is a photon
instead of a proton, the experimental data from L3 \ct{L300a} are larger than
the upper estimate $\textstyle{\frac{4}{5}}\si^{\rm Tot}_{\ga^*\ga}({Q^2=
4m_c^2})$ which in turn is larger than the model, showing approximately the same
relative magnitude as in the proton case.

So the charm data may indicate that $\ga\ga$ and $\ga^*\ga$ processes 
are really rather different from the corresponding $\ga p$ reactions. 
If treated in isolation, the $\ga\ga$ and $\ga^*\ga$ data can be
described by

$\bullet$ increasing the fraction of the hard-pomeron and decreasing the 
fraction of the soft pomeron in $\si^{\rm Tot}_{\ga\ga}$ and, possibly, 
giving the hard pomeron a stronger energy dependence than we have used 
in our model. 

$\bullet$ making corresponding changes in $F_2^\ga$ and modifiying the $Q^2$
dependence of both terms, that is discarding the simple picture of 
(\ref{dlpga}).These modifications can not be excluded by the
present data.

Note that our calculation does not include central production of charmed
quark-anti\-quark pairs (doubly-resolved photons in pQCD language), but
for this contribution to have a significant effect it would need to
play a more important role in $\gamma$-$\gamma$ collisions than in 
$\gamma$-proton interactions. 
 
We note that results from perturbative QCD 
\cite{FKL00} using the photon structure function from \cite {GRS99} report 
no such discrepancy with the data. This is interesting as for $\gamma^* p$
reactions the results of perturbative QCD are essentially indistinguishable
from our model. The comparison between perturbative QCD and the $\ga\ga$ 
charm data was done for the full process $e^+e^- \to e^+e^- c \bar c$, so 
we have converted our $\ga\ga$ cross section to the full $e^+e^-$ cross 
section using the equivalent photon approximation \cite{BGMS75}. The 
threshold was taken as $z_{th}=4 m_c^2$ and $Q^2_{max}= 4 m_c^2$, 
in the notation of \ct{PDG00}. The result for
$m_c$=1.25 is shown in figure \rf{epem1}(a) and compared with
results from pQCD and experimental data. Here the agreement with the
perturbative QCD results is rather good and the discrepancy with experiment
seems not  dramatic. One reason for this is that the box diagram remains
important up to the highest values of $\sqrt{s_{e^+e^-}}$, see figure
\rf{epem1}(b), and the discrepancy at high $W$ in figure \rf{sigarech} is
smeared out in the full $e^+e^-$ data and is much less visible. This
emphasises the importance of comparing models with $\si_{\ga^*\ga^*}$
rather than with $\si_{e^+e^-}$. We also note
that there  are significant differences at present among the preliminary data
sets available, so it is perhaps too early to attempt to draw significant 
conclusions.

\begin{figure}
\leavevmode
\begin{center}
\begin{minipage}{9cm}
\epsfxsize9cm
\epsffile{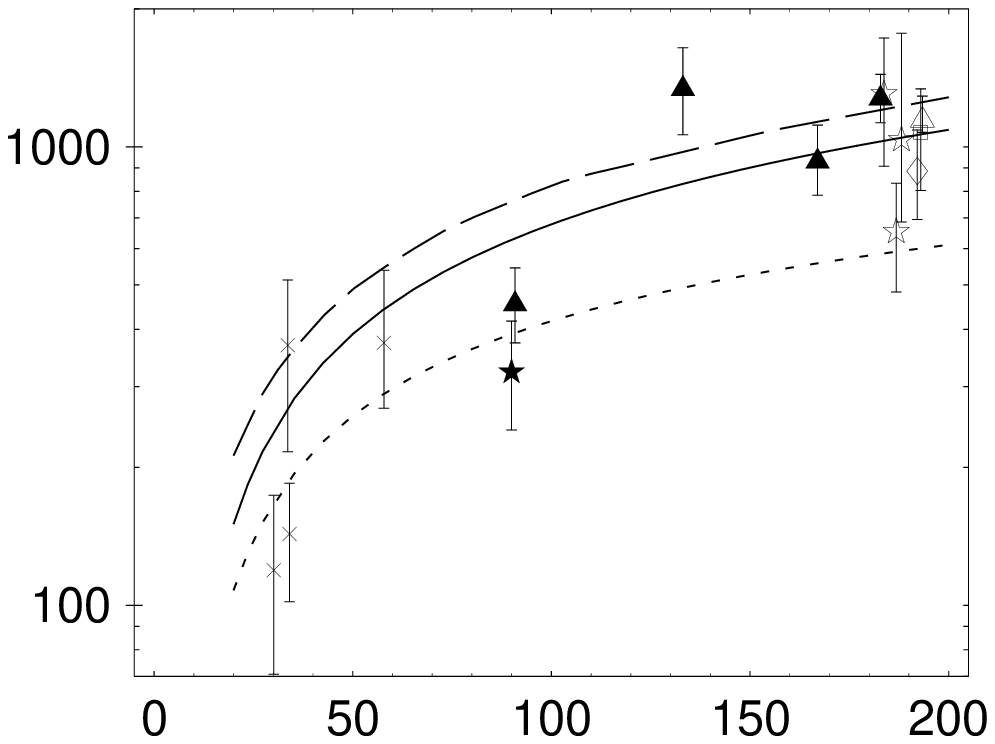}
\begin{picture}(0,0)
\setlength{\unitlength}{1.5mm}
\put(3,46){$\si$}
\put(2,43){[pb]}
\put(35,0){$\sqrt{s_{ee}}$ [GeV]}
\end{picture}
\end{minipage}
\hspace{10mm}
\hfill
\end{center}
\caption{Cross section for the reaction $e^+e^- \to e^+e^- c \bar c$. 
The solid line is our model with $m_c$ =1.25, the dotted line the box diagram
alone  and the long dashed line is  NLO perturbative QCD \protect\ct{FKL00}
with  $m_c$ =1.3 GeV. The data are: L3\protect\cite{L300a}, triangles;
ALEPH\protect\cite{ALEPHcc}, stars;
DELPHI \protect\cite{DELPHIcc}, diamonds; 
OPAL \protect\cite{OPAL99a}, boxes. The results at lower energies (crosses) are
from \protect\cite{JADEcc,TPCcc,TOPAZcc,AMYcc}.
Open symbols refer to 
preliminary results. 
\lb{epem1}} \end{figure}

Although it is perhaps premature to draw firm conclusions from the
charm data, it is clear that our nonperturbative model cannot give the full
answer when both photons have high virtuality since it decreases much faster
with increasing virtuality than purely perturbative contributions.  The
dipole-dipole cross section (\ref{sidip}) behaves for small values of 
$R_1=R_2=R$ as $R^4$, and therefore the $\ga^*\,\ga^*$ cross section  
decreases like $1/Q^4$ for $Q^2=Q_1^2=Q_2^2$  and fixed $Q^2/W^2$. 
The perturbative contributions decrease, up to logarithms, like
$1/Q^2$ \cite{BHS97}. This follows from a simple dimensional
argument. In perturbation theory with massless quarks, in forward 
scattering the only dimensioned quantities are $W^2$ and $Q^2$ as
no internal scale appears. Therefore for fixed $Q^2/W^2$ the cross 
section has to be proportional to $1/Q^2$. 
\begin{figure}
\leavevmode
\begin{center}
\begin{minipage}{60mm}
\epsfxsize60mm
\epsffile{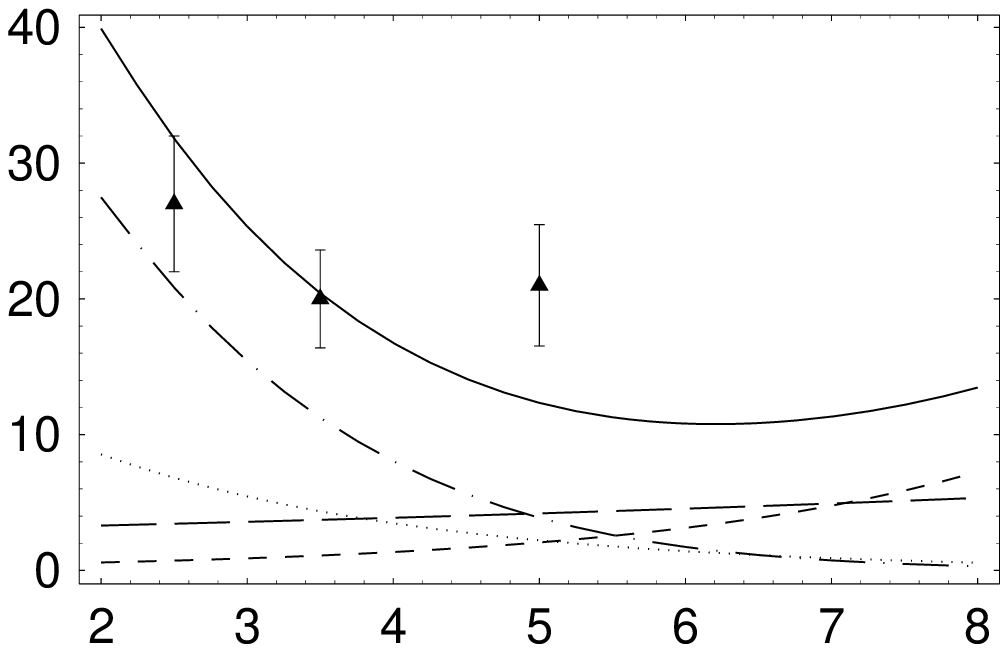}
\begin{picture}(0,0)
\setlength{\unitlength}{1.2mm}
\put(25,2){\small $Y$}
\put(-6,30){$\si$}
\put(-8,25){[nb]}
\put(20,-3){(a)}
\end{picture}
\end{minipage}
\begin{minipage}{60mm}
\epsfxsize60mm
\epsffile{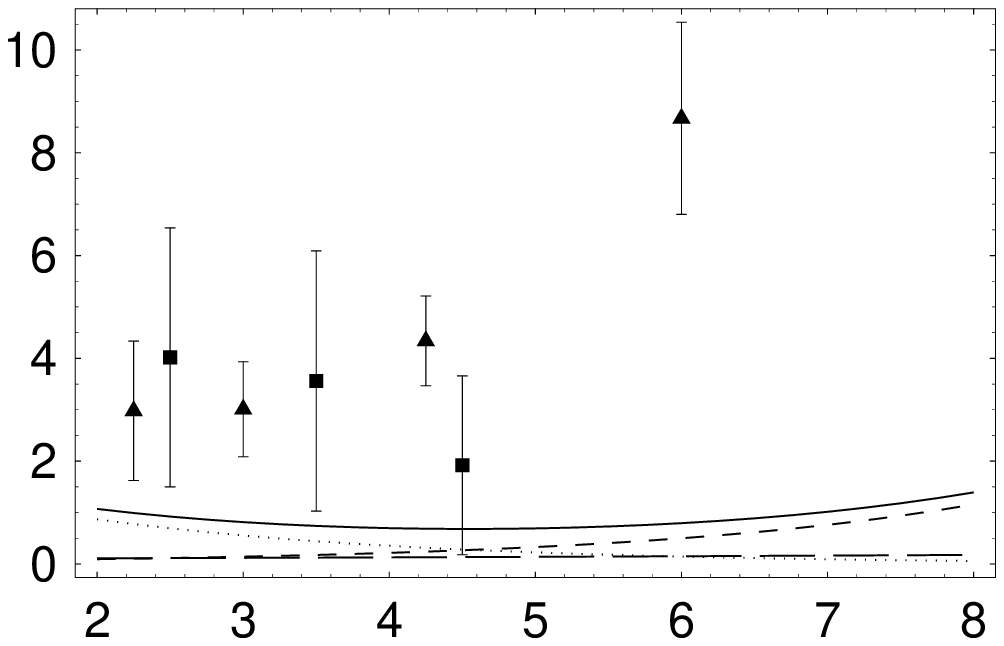}
\begin{picture}(0,0)
\setlength{\unitlength}{1.2mm}
\put(25,2){\small $Y$}
\put(20,-3){(b)}
\end{picture}
\end{minipage}
\end{center}
\caption{$\si_{\ga^*\ga^*}$ with box graph contribution subtracted. 
(a) L3 data \ct{L3star} at $\langle Q^2 \rangle = 3.5$ GeV$^2$. (b) 
preliminary L3 data \ct{L3starn} at $\langle Q^2 \rangle = 15$ GeV$^2$, 
 triangles and preliminary OPAL data \ct{OPALstar} at $\langle Q^2 \rangle  
= 17$ GeV$^2$, boxes.
The separate theoretical contributions are: long dashes, soft pomeron;
short dashes, hard pomeron; dots,
reggeon; solid, sum.} \lb{GSTAR}\end{figure}
\begin{figure} \leavevmode \begin{center} \begin{minipage}{50mm}
\epsfxsize50mm
\epsffile{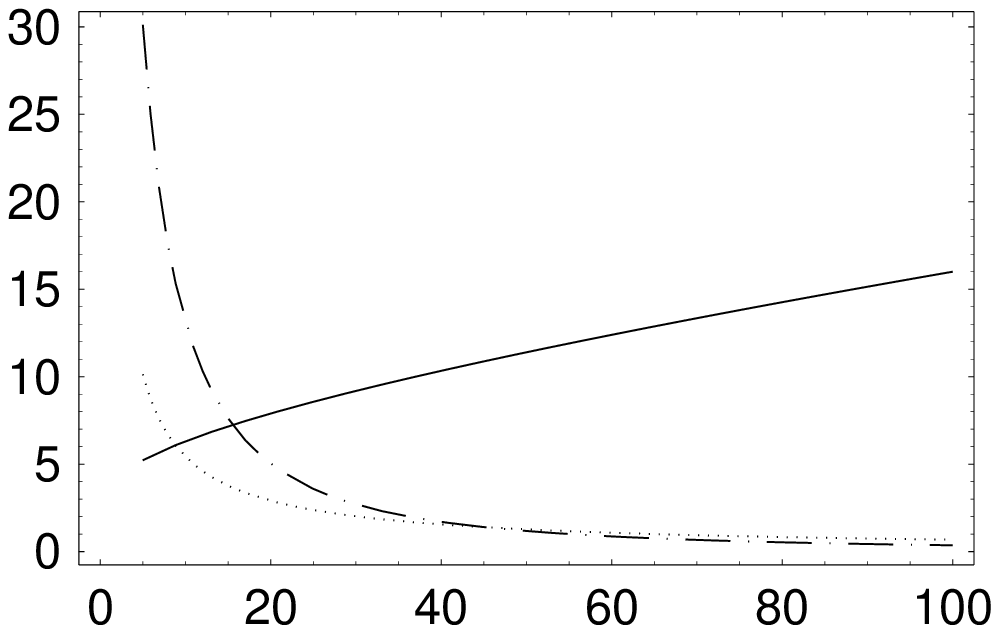}
\hspace{12mm}
\begin{picture}(0,0)
\setlength{\unitlength}{1mm}
\put(-6,30){$\si$}
\put(-8,25){[nb]}
\put(28,30){3 GeV$^2$}
\put(25,2){\small $W$ [GeV]}
\end{picture}
\end{minipage}
\begin{minipage}{50mm}
\epsfxsize50mm
\epsffile{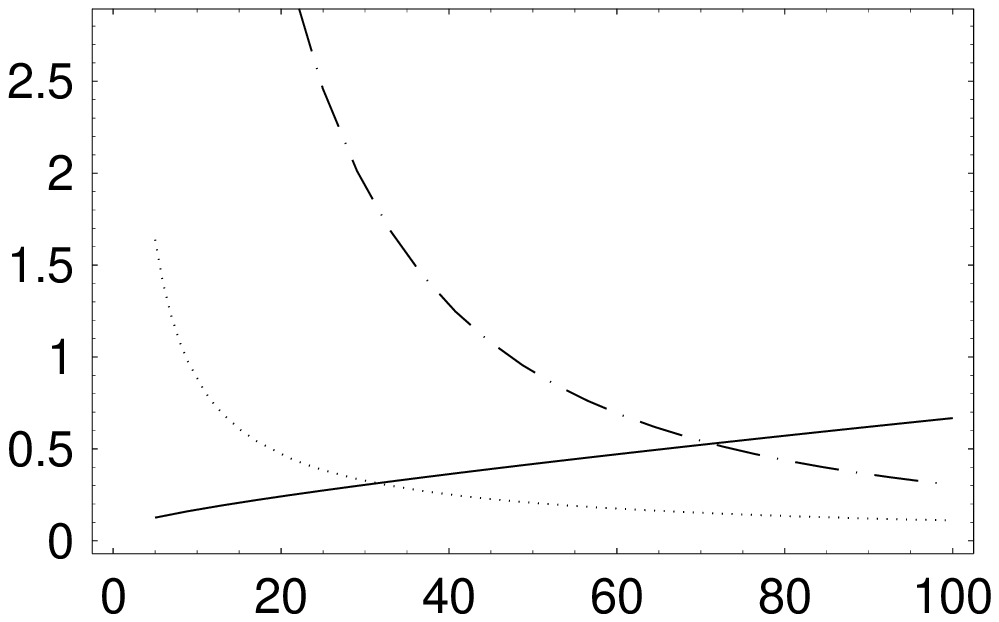}
\begin{picture}(0,0)
\setlength{\unitlength}{1mm}
\put(28,30){17 GeV$^2$}
\put(25,2){\small $W$ [GeV]}
\end{picture}
\end{minipage}
\end{center}
\caption{Separate contributions to the $\si_{\ga^*\ga^*}(W)$ at fixed values of
$Q_1^2=Q_2^2 = Q^2$ as indicated in the figures: 
solid, pomerons; dot-dashes, fixed pole (box diagram); dots, reggeon.}
\lb{gastgast1} \end{figure}
\begin{figure}
\leavevmode
\begin{center}
\begin{minipage}{50mm}
\epsfxsize50mm
\epsffile{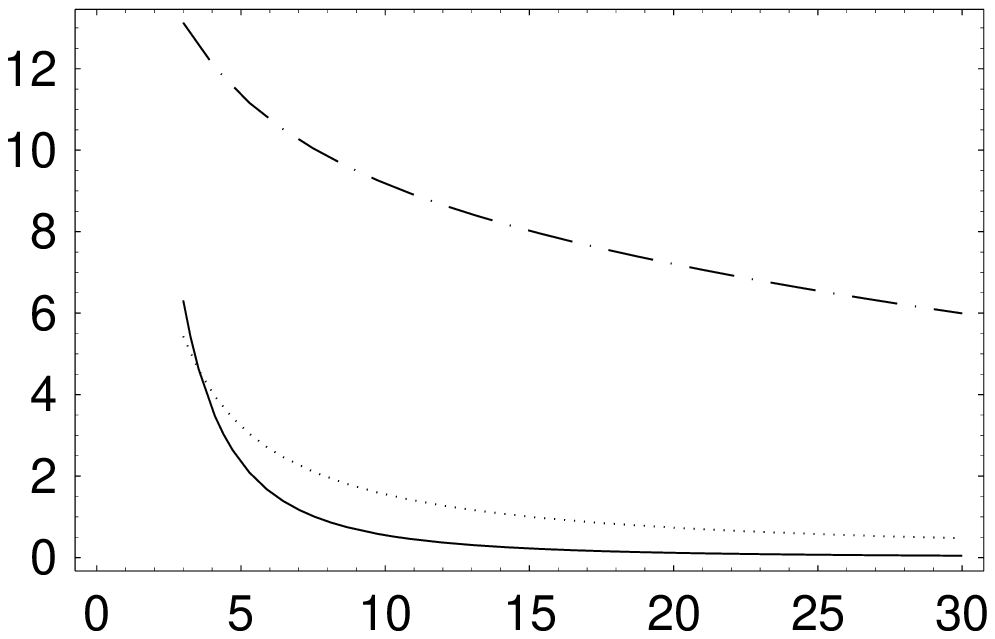}
\begin{picture}(0,0)
\setlength{\unitlength}{1mm}
\put(28,30){10 GeV}
\put(25,1){\small $Q^2$ [GeV$^2$]}
\put(-6,30){$\si$}
\put(-8,25){[nb]}
\end{picture}
\end{minipage}
\begin{minipage}{50mm}
\epsfxsize50mm
\epsffile{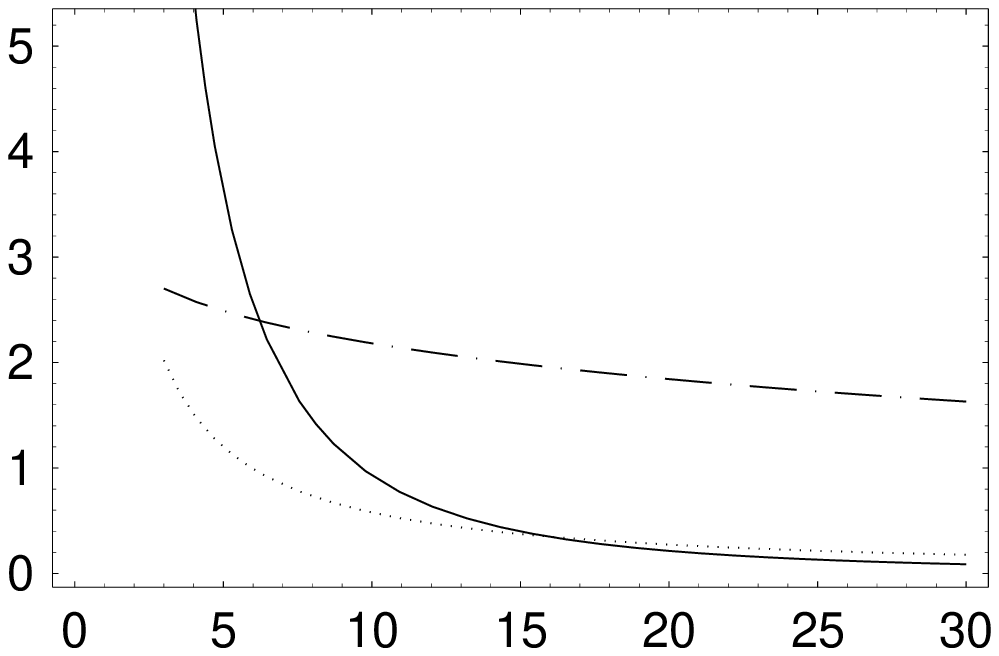}
\begin{picture}(0,0)
\setlength{\unitlength}{1mm}
\put(28,30){30 GeV}
\put(25,1){\small $Q^2$ [GeV$^2$]}
\end{picture}
\end{minipage}
\end{center}
\caption{Separate contributions to the $\si_{\ga^*\ga^*}(Q^2)$ at fixed values
of $W$ as indicated in the figures:  solid, pomerons;  dot-dashes, fixed pole 
(box diagram); dots, reggeon.}\lb{gastgast2}
\end{figure}
However our model can be used as an
estimate for the non-perturbative background in $\ga^*\ga^*$ 
reactions. In figures \rf{gastgast1} and \rf{gastgast2} we show the 
different contributions 
to $\si^{\rm Tot}_{\ga^*\ga^*}(Q^2)$ as a function of the 
common virtuality $Q^2$ at fixed $W$ and as a function of $W$ at fixed $Q^2$.
In figure \ref{GSTAR} the theoretical contributions are displayed as a function
of $Y\approx \ln(W^2/Q^2)$ for fixed $\langle Q^2 \rangle =
3.5$ and 16 GeV$^2$. They are compared with the   L3
\cite{L3star,L3starn} and OPAL \cite{OPALstar} data at $\langle Q^2 \rangle =
3.5$ and $\langle Q^2 \rangle = 15$ and 17  GeV$^2$ after subtraction of the 
box diagram. We note that there is also a very clear purely-nonperturbative
signal visible beyond the box diagram. 

\subsection{Summary and Conclusions}

We have used a simple dipole-dipole approach, adopted to the two pomeron
picture \ct{DL98} in order to describe a great variety of high energy 
reactions.
The picture of dipole-dipole scattering is a consequence of our nonperturbative
approach which starts from the evaluation of lightlike Wilson
loops\ct{Nac91,KD91}. The dependence of the scattering amplitude on the dipole
sizes is also determined in our model and related to low-energy and lattice
results \ct{Dos87,DS88,DFK94}. The energy dependence is put into the model by
hand and inspired by the $Q^2$ dependence of the hard and soft contribution
found in \ct{DL98}: if at least one dipole is smaller than a critical value
$R_c$ the energy dependence is governed by the hard pomeron, otherwise by the
soft one. The numerical value $R_c\approx 0.22$ fm was adjusted from comparison
with the proton structure function. With this single parameter the $Q^2$
dependence of the soft and hard contribution of the proton structure
function obtained in \ct{DL98} is reproduced very
well.  The behaviour of different reactions is then  controlled solely by the
different  wave function of the participating photons and  hadrons. We have
introduced no saturation mechanism into the dipole cross section and the
analysis shows that there is no compelling reason to do so. 

There is one reaction which seems to jeopardize our simple picture, namely
charm production in photon-photon reactions where our model underestimates the
results from L3 \ct{L300a} for the reaction $\gamma\gamma \to c \bar c X$ by
about a factor of two. We have discussed this question in detail also in an
less model-dependent approach and argued that, interesting as the discrepancy
is, it might be to premature to draw final conclusions. At any rate future data
for this reaction might be very important for our general understanding of the
underlying mechanisms of the dipole approach. 

\begin{table}\begin{center}
\begin{tabular}{|c|c|c|c|}\hline
$W$&$pp$&$\gamma p$&$\ga\ga$\\ 
$[$GeV$]$&&&\\
\hline
20&0.0023&0.037&0.071\\
100&0.007&0.10&0.19\\
200&0.011&0.16&0.27\\
1800&0.048&0.45&0.62\\
\hline
\end{tabular}
\end{center}
\caption{The ratio of the hard pomeron to the total 
pomeron contribution for the total cross sections of different 
reactions; without any unitarity corrections}
\lb{softhard}\end{table}

A consequence of our approach is that the hard pomeron
is not a product of  perturbative evolution but is also present in soft
processes. For example, we  have seen that the $\gamma\gamma$ cross section
receives a non-negligible hard contribution due to the pointlike coupling of
the photon and the resulting  singularity of the photon wave function at the
origin. Similarly there is a hard contribution to the $\ga p$ cross section. It
is compatible with  but not demanded by experiment. There is necessarily also a
non-zero hard contribution to proton proton scattering. In table \rf{softhard}
we
give the ratio of  the hard to the soft plus hard contribution for different
reactions and centre-of-mass energies. As can be seen the hard component 
in
proton-proton scattering  is so small as to be within the limits of
experimental error at present energies. At higher energies, where its presence
might be expected to be observable, it will be suppressed by unitarity
corrections, therefore the values for proton-proton scattering represent an
upper limit. As the proton, unlike the photon, is a genuinely nonperturbative
object, the difficulty of detecting a hard contribution in proton proton
scattering means that the $\gamma p$ total cross section is of considerable
importance in this  respect.

\vskip 2truecm

Acknowledgements: We acknowledge helpful discussion with O. Nachtmann, 
H. Pirner, A. Shoshi, F. Steffen. The work was supported in part by 
PPARC grant PPA/G/0/1998 and by BMBF grant 05HT1VHA/0.

\clearpage

\end{document}